\documentclass[journal=jctc,manuscript=article]{achemso}

\usepackage[version=3]{mhchem} 
\usepackage[final]{changes}
\usepackage{todonotes}
\usepackage{mciteplus}
\usepackage{graphicx}
\usepackage{subcaption}
\usepackage{geometry}
\usepackage{caption}
\usepackage{booktabs}
\usepackage{siunitx}
\usepackage{xurl}
\author{Tianyi Li}
\affiliation[1]
{Center on Frontiers of Computing Studies, School of Computer Science, Peking University, Beijing 100871, China}
\alsoaffiliation[2]
{École Normale Supérieure Paris-Saclay, Université Paris-Saclay, 91190 Gif-sur-Yvette, France}
\author{Yumeng Zeng}
\affiliation[1]
{Center on Frontiers of Computing Studies, School of Computer Science, Peking University, Beijing 100871, China}
\author{Qiming Ding}
\affiliation[1]
{Center on Frontiers of Computing Studies, School of Computer Science, Peking University, Beijing 100871, China}
\author{Zixuan Huo}
\affiliation[1]
{Center on Frontiers of Computing Studies, School of Computer Science, Peking University, Beijing 100871, China}
\author{Xiaosi Xu}
\affiliation[3]
{Graduate School of China Academy of Engineering Physics, Beijing 100193, China}
\author{Jiajun Ren}
\affiliation[4]
{Key Laboratory of Theoretical and Computational Photochemistry, Ministry of Education, College of Chemistry, Beijing Normal University, Beijing 100875, China}
\author{Diandong Tang}
\email{tangdd@uw.edu}
\affiliation[5]
{Department of Chemistry, University of Washington, Seattle, WA 98195-1700, USA}
\author{Xiaoxia Cai}
\email{xxcai@ihep.ac.cn}
\affiliation[6]
{Institute of High Energy Physics, Chinese Academy of Sciences, Beijing 100049, China}
\author{Xiao Yuan}
\email{xiaoyuan@pku.edu.cn}
\affiliation[1]
{Center on Frontiers of Computing Studies, School of Computer Science, Peking University, Beijing 100871, China}

\title
  {\replaced{Efficient Quantum Simulation of Non-Adiabatic Molecular Dynamics with Precise Electronic Structure}{Efficient Simulation of Non-adiabatic Molecular Dynamics using Adaptable Quantum Algorithms}}

\abbreviations{}

\begin{document}
\begin{tocentry}

\includegraphics[width=8cm, height=4cm, keepaspectratio]{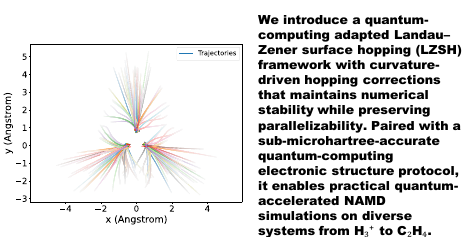}
\end{tocentry}

\begin{abstract}

In the study of non-adiabatic chemical processes such as photocatalysis and photosynthesis, non-adiabatic molecular dynamics (NAMD) is an indispensable theoretical tool, which requires precise potential energy surfaces (PESs) of ground and excited states. Quantum computing offers promising potential for calculating \replaced{PESs that are intractable for classical computers}{classically intractable PESs}\comment{R3 Q3}. However, its realistic application poses significant challenges to the development of quantum algorithms that are sufficiently general to enable efficient and precise PES calculations across chemical systems with diverse properties, as well as to seamlessly adapt existing NAMD theories to quantum computing. In this work, we introduce a quantum-adapted extension to the Landau-Zener-Surface-Hopping (LZSH) NAMD. This extension incorporates curvature-driven hopping corrections that protect the population evolution while maintaining the efficiency gained from avoiding the computation of non-adiabatic couplings (NACs), as well as preserving the trajectory independence that enables parallelization. Furthermore, to ensure the high-precision PESs required for surface hopping dynamics, we develop a sub-microhartree-accurate PES calculation protocol. This protocol supports active space selection, enables parallel acceleration either on quantum or classical clusters, and demonstrates adaptability to diverse chemical systems—including the charged H$_3^+$ ion and the C$_2$H$_4$ molecule, a \replaced{prototypical}{classical}\comment{R3 Q3} multi-reference benchmark. This work paves the way for practical application of quantum computing in NAMD, showcasing the potential of parallel simulation on quantum-classical heterogeneous clusters for ab-initio computational chemistry.

\end{abstract}

\section{Introduction}

Ab initio molecular dynamics (AIMD) simulations are indispensable for elucidating mechanisms underlying chemical and biological processes, providing atomistic insights into phenomena ranging from charge carriers dynamics in materials\cite{Prezhdo2023}, excited-state dynamics of transition metal complexes\cite{Gonzalez2021}, proton transfer in solvation process\cite{Li2021}, to photochemical reactions \cite{curchod2018ab, mai2020molecular}. A foundational early approach of AIMD is the Born-Oppenheimer molecular dynamics\deleted{(BOMD)}\comment{R2Q1} framework, which leverages the Born-Oppenheimer\deleted{(BO)}\comment{R2Q1} approximation to decouple electronic and nuclear degrees of freedom. This approach has been proven valuable for simulating equilibrium properties and slow dynamical processes in systems where the \replaced{Born-Oppenheimer}{(BO)}\comment{R2Q1}  approximation holds\cite{born1927, barnett1993, kuhne2007, niklasson2008}.

However, the \replaced{Born-Oppenheimer}{(BO)}\comment{R2Q1}  approximation breaks down when the energy gap between electronic states becomes close, leading to strong non-adiabatic effects, which is crucial for understanding a broad spectrum of chemical phenomena. This occurs in scenarios involving conical intersections, avoided crossings, or ultra-fast electronic transitions, such as in photochemistry\cite{Chen2004,Curchod2020,Romero2014}, charge transfer\cite{Weinberg2012,Huynh2007,Hammes-Schiffer2010}, or vibronic relaxation processes\cite{Domcke2004, Worth2004, Levine2007, Matsika2007, Matsika2011, Levine2019, Shen2020, Matsika2021}. Under these conditions, \replaced{Born-Oppenheimer molecular dynamics}{BOMD}\comment{R2Q1} fails to capture the traveling of nuclear wavepackets across multiple \added{potential energy surfaces}(PESs)\comment{R2 Q1}, and is not able to describe excited-state dynamics and simulate photophysics and photochemistry reactions.

 \replaced{To fully simulate those processes, non-adiabatic molecular dynamics (NAMD) is necessary. Full quantum dynamics treats both electronic and nuclear degrees of freedom quantum mechanically, with the multi-configurational time-dependent Hartree\cite{BECK20001} method being a prominent example; however, its computational cost escalates rapidly with system size. In contrast, mixed quantum-classical dynamics approximates nuclear motion classically while retaining quantum-mechanical treatment of electrons, enabling efficient simulations of larger systems over relevant timescales. Several schemes fall under this category, including mean-field approaches, ab initio multiple spawning, and trajectory surface hopping\cite{review_big,Li2005,BenNun2000,Preston1971}. Among surface hopping methods, the widely adopted fewest switches surface hopping (FSSH)\cite{Tully1990} propagates nuclear trajectories classically on a single active PES while allowing stochastic hops between states based on transition probabilities. These probabilities are computed equally from three key quantities: the non-adiabatic coupling (NAC), which represents the interaction between electronic states induced by nuclear motion; the nuclear time step, which scales the probability to ensure proper integration over the trajectory; and the electronic coefficients, which encode the quantum amplitudes and coherences among states—though the update of these electronic coefficients itself relies on the NAC to capture non-adiabatic effects during propagation. In addition to NAC, energy gaps between PESs and their derivatives offer more accessible electronic properties for driving state transitions, leading to efficient variants such as \added{Landau-Zener-Surface-Hopping}\comment{R2 Q1} (LZSH)\cite{Belyaev2014}, Zhu-Nakamura surface hopping (ZNSH)\cite{zhu1994,zhu2001,zhu2002}, and curvature-driven surface hopping ($\kappa$SH)\cite{shu2022,zhao2023direct,zhao2023nonadiabatic}. These protocols construct hopping events without explicit computation of NAC, making them well-suited for interfacing with electronic structure methods that may not readily provide such quantities\cite{Suchan2021}.}{To fully simulate those processes, full quantum dynamics with both electronic and nuclei degrees of freedom has been developed, such as the multi-configurational time-dependent Hartree (MCTDH) method, but the computation cost quickly builds up, making non-adiabatic molecular dynamics (NAMD) necessary. Several schemes have been developed for NAMD, including mean field approaches, ab initio multiple spawning, and trajectory surface hopping\cite{review_big,Li2005,BenNun2000,Preston1971}. Surface hopping methods, such as the widely adopted fewest switches surface hopping (FSSH)\cite{Tully1990}, propagate nuclear trajectories classically on a single active PES while allowing stochastic hops between states based transition probabilities computed by NAC. In addition to NAC, energy gaps between PESs and their derivatives serve as better accessible electronic property in NAMD, also provide insights in state transitions, and lead to efficient variants such as LZSH\cite{Belyaev2014}, Zhu-Nakamura surface hopping (ZNSH)\cite{zhu1994,zhu2001,zhu2002}, and curvature-driven surface hopping ($\kappa$SH)\cite{shu2022,zhao2023direct,zhao2023nonadiabatic}. These protocols construct hopping events without the need for explicit computation of time-derivative non-adiabatic couplings, making it well-suited for interfacing with electronic structure methods that may not readily provide such quantities, while still enabling simulations of large systems over relevant timescale\cite{Suchan2021}. 
}\comment{R2Q1,2,3,4; R3 Q3}

Despite these advances, solving the multi-state electronic structure remains a major challenge for NAMD on classical computers. Accurate treatment of conical intersections and strongly correlated systems often requires full configuration interaction (FCI), which scales exponentially with the number of orbitals and electrons, necessitating an exponential number of Slater determinants\cite{Eriksen2021}, which is intractable for classical computer. While density functional theory (DFT) provides a computationally efficient alternative with $O(N^3)$ or $O(N^4)$ scaling, it faces challenges in accurately describing multi-configurational wavefunctions and excitation energies in non-adiabatic regimes\cite{Cohen2012,Lacombe2023}. These challenges motivate the exploration of quantum algorithms, which inherently exploit superposition and entanglement to efficiently solve the electronic Schrödinger equation for correlated systems, promising exponential speedups for NAMD in regimes inaccessible to classical\added{-computing}\comment{R3 Q3} methods\cite{McArdle2020,HeadMarsden2021,Lee2023}.

Quantum algorithms for chemistry simulation are exemplified by the quantum phase estimation (QPE) method in the fault-tolerant quantum computing (FTQC) regime, which can theoretically achieve accuracy comparable to FCI, provided a suitable initial state is prepared (e.g., via adiabatic state preparation)\cite{aspuru2005simulated,reiher2017elucidating}. Nevertheless, current quantum devices fall short in supporting QPE circuits of practical width, depth, qubit fidelity, and gate fidelity, compounded by the immaturity of quantum error correction. Near-term quantum algorithms, represented by \replaced{Variational Quantum Eigensolver (VQE)}{VQE}\comment{R3 Q2}, offer a compromise by adopting a heuristic time complexity and tolerating moderate noise levels\cite{peruzzo2014variational,mcclean2016theory}. Grounded in the variational principle, VQE optimizes parametrized quantum circuits to approximate molecular energy spectra and corresponding electronic states. For larger-scale systems, the second-quantization on which VQE framework based enables the selection of chemically significant molecular orbitals (MOs), forming an active space that captures essential chemical properties within constrained circuit sizes\cite{reiher2017elucidating,cailaoshi}. This complete active space (CAS) method is particularly vital for extending quantum\added{-computing}\comment{R3 Q3} simulations to realistic molecular systems without exceeding resources of noisy intermediate-scale quantum (NISQ)\added{\cite{nisqPreskill_2018}}.\comment{R3 Q5}

Advancements in NISQ algorithms have significantly enhanced the computation of molecular excited states. Among these, the variational quantum deflation (VQD) method extends the VQE by incorporating overlap penalties to enforce approximate orthogonality with previously computed states, enabling sequential search of higher-energy eigenstates\cite{vqd1,conintersection_water}. However, its iterative framework escalates computational demands, limits parallelization, and propagates errors in noisy settings. In contrast, subspace-based approaches enhance efficiency by restricting calculations to predefined excitation sectors. For instance, the subspace-search VQE (SSVQE) simultaneously optimizes a set of orthogonal initial states, reducing optimization iterations comparing to serial search\cite{nakanishi2019subspace}. Complementing subspace concepts with sampling strategy, the sample-based quantum diagonalization (SQD) constructs and diagonalizes effective Hamiltonian matrices via quantum sampling, claimed to support large qubit systems and mitigating noise effects, yet it relies heavily on ample sampling for achieving high precision\cite{sqd2,sqd_1}. Building on similar foundations, the \replaced{Quantum Subspace Expansion (QSE)}{QSE}\comment{R2 Q1} employs explicit projections onto Fermionic subspaces, yielding physically guaranteed variational upper bounds for excited-state energies\cite{McClean2017,Tong2022,Takeshita_2020,Urbanek_2020}. Another prominent approach, the quantum equation-of-motion (QEOM), leverages subspace concepts while incorporating\deleted{classical} equation-of-motion (EOM) formalisms \added{in classical-computing quantum chemistry}\comment{R3 Q3}, could ensure size-intensivity and inherently introduce the contribution of de-excitation\cite{compareqseeom2,ollitrault2020quantum,Grimsley2025}. However, for QSE and QEOM, the choice and number of operators used in subspace construction determines the algorithm's ability to capture physical properties\cite{jctc1} as well as its efficiency and robustness. The exquisite design of subspaces tailored to specific systems remains a on-going topic. Furthermore, to meet the NISQ constraints, optimization strategies at the implementation level leaves a valuable area for exploration.

\replaced{In this work, we introduce a practical quantum-computing NAMD framework, seamlessly integrated with our sub-microhartree-accuracy calculations of PESs across diverse chemical systems, including H$_3^+$ and C$_2$H$_4$, and compatible with parallel acceleration on both 
real quantum computers and quantum algorithm simulator on classical computers. While recent explorations \citep{jctc1,chemical_science} have advanced quantum-computing electronic structure solvers by comparing methods such as QEOM, QSE, and de-excited QSE, or employing VQD for sequential orthogonal excited-state searches, our approach pursues a distinct goal of achieving sub-microhartree precision through a hybrid subspace-based quantum-computing electronic structure solver that adapts operator selections to different chmicals, enhancing accuracy via problem-adapted subspace operator selection and integration of SSVQE for parallel optimization of multiple reference states prior to QSE application. We also rigorously assess the numerical stability of quantum-computed PESs and incorporate an efficient curvature-driven correction scheme for state transitions tailored to quantum-computing electronic structure solvers. In the NAMD evolution, whereas prior works focus on FSSH requiring NAC computations, we focus on adapting LZSH for quantum algorithms, introducing an improved LZSH scheme that stabilizes dynamics more efficiently. These innovations, combined with a two-level parallelization framework and task-specific algorithmic extensions, yield substantial speedups without compromising PES precision, collectively elevating the robustness, efficiency, and practical viability of quantum-enhanced LZSH-NAMD simulations for broader chemical research.}{In this work, we introduce a practical quantum-computing NAMD framework, seamless compact with our sub-microhartree-accuracy calculation of PESs across diverse chemical systems including H$_3^+$ and C$_2$H$_4$, with compatibility for parallel acceleration on both classical simulators and quantum processors. Comparing to recent explorations \citep{jctc1,chemical_science}, we present a hybrid subspace-based quantum electronic structure solver that is problem-adaptable as well, and rigorously assess the numerical stability of quantum-computed PESs, incorporate an efficient curvature-driven correction scheme of state transition tailored for quantum solvers. These innovations collectively bolster the robustness and enhance the overall practicality of quantum-enhanced LZSH-NAMD simulations which preserve efficiency. Moreover, to further optimize the implementation, we introduce a two-level parallelization framework alongside task-specific algorithmic extensions, yielding substantial speedups without compromising PES precision. Taken together, these advances markedly elevate the practical viability of quantum-computing NAMD for broader chemical research.}\comment{R2 Q22, R3 Q3}

\section{Method}

\subsection{Capturing Molecular Properties on Quantum Computer}
\begin{figure}
    \centering
    \includegraphics[width=0.98\linewidth]{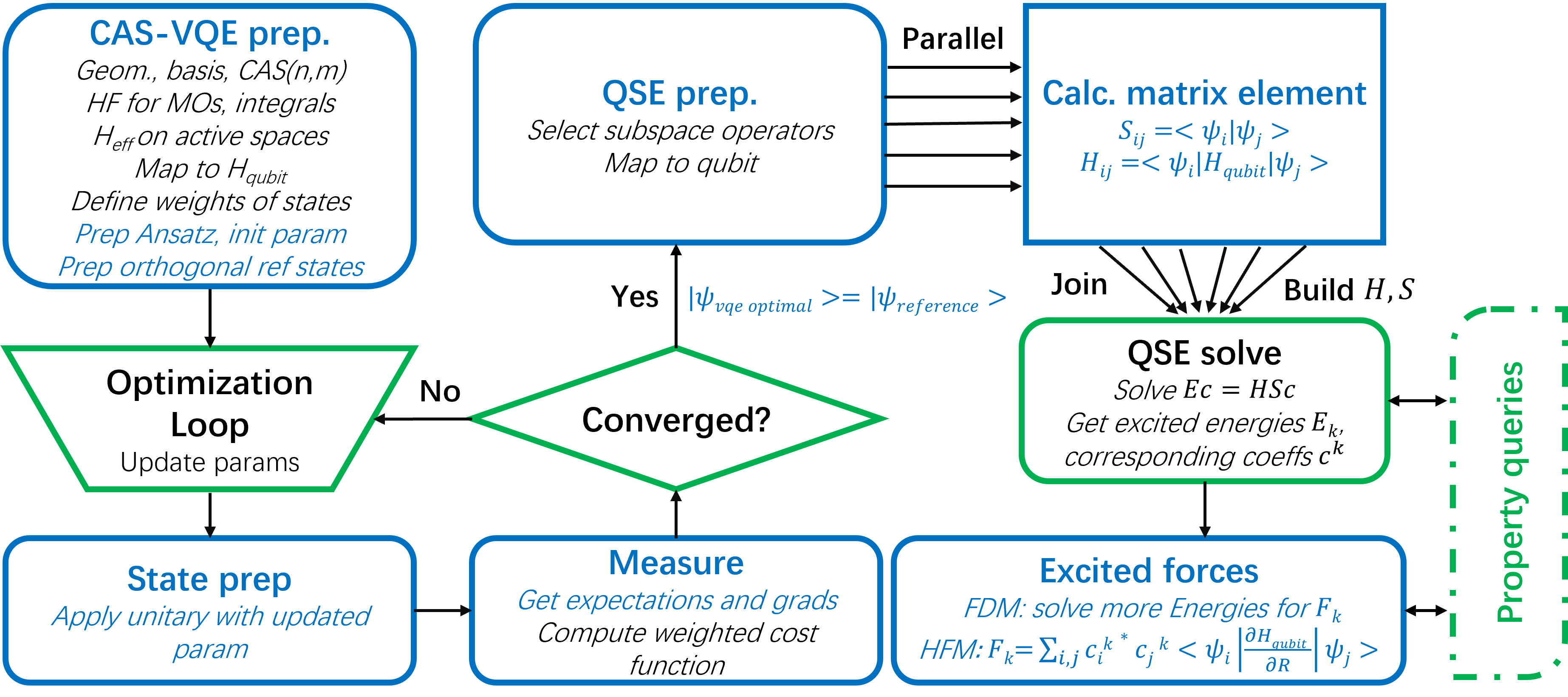}
    \caption{Work flow of on-the-fly quantum solver for electronic-structure observables, with parallization of QSE.}
    \label{fig:placeholder}
\end{figure}
\subsubsection{Solving Electronic Structure in Active Spaces}
\replaced{To simulate molecular electronic structure on a quantum computer with flexibility, we employ the second-quantized representation of the molecular Hamiltonian within a selected complete active space (CAS)~\cite{Hempel_2018}. In quantum chemistry, CAS classifies molecular orbitals into core (always doubly occupied), active (partially occupied), and virtual (always unoccupied) sets, generating a wavefunction as a configuration interaction expansion within the active orbitals, which resolve the electron correlation problem in strongly correlated systems such as bond dissociation or transition metal complexes.}{To simulate molecular electronic structure on a quantum computer with flexibility, we employ the second-quantized representation of the molecular Hamiltonian within a selected active space\cite{Hempel_2018}. This approach reduces the resource cost by focusing on a subset of orbitals that capture the essential electronic correlations, while treating the remaining orbitals at a mean-field level.}\comment{R2 Q5}

The electronic molecular Hamiltonian in second quantization is expressed as\cite{szabo1989modern}
\begin{equation}
\hat{H} = \sum_{pq} h_{pq} \hat{a}_p^\dagger \hat{a}_q + \frac{1}{2} \sum_{pqrs} g_{pqrs} \hat{a}_p^\dagger \hat{a}_q^\dagger \hat{a}_r \hat{a}_s,
\label{eq:hamiltonian_second_quant}
\end{equation}

\replaced{where $\hat{a}_p^\dagger$ and $\hat{a}_p$ are the fermionic creation and annihilation operators for orbital $p$, $h_{pq}$ are the one-electron integrals, and $g_{pqrs}$ are the two-electron integrals. These integrals are obtained using classical-computing quantum chemistry software, typically through Hartree-Fock calculations. The indices $p, q, r, s$ run over the MOs.}{where \(\hat{a}_p^\dagger\) and \(\hat{a}_p\) are the fermionic creation and annihilation operators for orbital \(p\), \(h_{pq}\) are the one-electron integrals, and \(g_{pqrs}\) are the two-electron integrals. The indices \(p, q, r, s\) run over the MOs.}\comment{R2 Q7;R3 Q3}

In the CAS framework\cite{Tilly2021}, we partition the orbitals into core (inactive), active, and virtual (inactive) sets. The core orbitals are doubly occupied, and their contributions are incorporated into an effective one-electron potential. The active space consists of \(m\) orbitals and \(n\) electrons, where strong correlations are expected, such as in bond-breaking regions or excited states. \added{Selecting the active orbitals is crucial for accuracy and efficiency. General strategies include identifying orbitals based on chemical intuition, such as valence orbitals involved in bonding or antibonding interactions, or using orbital energies and occupancies from preliminary calculations. For smaller active spaces (e.g., CAS(2,2) or CAS(4,4)), manual selection is common, visualizing molecular orbitals from Hartree-Fock calculations to choose frontier orbitals such as highest occupied molecular orbital, lowest unoccupied molecular orbital, or those directly participating in the chemical process of interest, ensuring the space captures the dominant static correlation with minimal computational cost. For larger active spaces (e.g., CAS(10,10) or beyond), automated strategies are preferred to handle complexity, such as the ranked-orbital approach\cite{rankorb}, entropy-based selection from uncorrelated natural orbitals combined with Density matrix renormalization group\cite{Luo_2017}, or machine learning selection\cite{mlorb}, which systematically expand the space while maintaining convergence. In this work, we use a selection strategy based on orbital energies and occupancies from preliminary calculations, where the active space comprises contiguous frontier orbitals around the Fermi level to efficiently capture dominant static correlations.}\comment{R2 Q6;R3 Q4}

The Hamiltonian is then restricted to excitations within this active space, yielding
\begin{equation}
\hat{H}_{\text{AS}} = \sum_{pq \in \text{AS}} h_{pq}^{\text{eff}} \hat{a}_p^\dagger \hat{a}_q + \frac{1}{2} \sum_{pqrs \in \text{AS}} g_{pqrs} \hat{a}_p^\dagger \hat{a}_q^\dagger \hat{a}_r \hat{a}_s,
\label{eq:hamiltonian_as}
\end{equation}
where \(h_{pq}^{\text{eff}}\) includes corrections from the inactive orbitals.

\replaced{Since quantum computers operate on qubits rather than fermions, the fermionic Hamiltonian must be mapped to a qubit Hamiltonian to enable simulation via quantum circuits. To achieve this, we apply the Jordan--Wigner (JW) transformation~\cite{Jordan1928}.}{To map this fermionic Hamiltonian onto a qubit system suitable for quantum circuit, we apply the Jordan-Wigner (JW) transformation\cite{Jordan1928}.}\comment{R2 Q7} The JW mapping encodes fermionic operators into \added{$O(1)$ number of} Pauli strings as
\begin{equation}
\hat{a}_p^\dagger = \frac{1}{2} (X_p - i Y_p) \prod_{j=0}^{p-1} Z_j, \quad \hat{a}_p = \frac{1}{2} (X_p + i Y_p) \prod_{j=0}^{p-1} Z_j,
\label{eq:jw_transform}
\end{equation}
\added{where \(X_p, Y_p, Z_p\) are Pauli operators acting on qubit \(p\)}\comment{R3 Q6}, and the product of \(Z\) operators enforces Fermionic anticommutation relations. This transformation requires \(2m\) qubits for a spin-orbital basis (or \(m\) qubits with spin-symmetry adaptations). Substituting Eq.~\eqref{eq:jw_transform} into Eq.~\eqref{eq:hamiltonian_as} results in a qubit Hamiltonian
\begin{equation}
\hat{H}_{\text{qubit}} = \sum_k c_k \hat{P}_k,
\label{eq:hamiltonian_qubit}
\end{equation}
\replaced{where $\hat{P}_k$ are Pauli strings mapped from effective one-electron integral terms and two-electron integral terms, $c_k$ are coefficients added up by the coefficients of the same Pauli strings.}{where $\hat{P}_k$ are Pauli strings and $c_k$ are coefficients derived from the integrals.}\comment{R2 Q8;R3 Q6}

\subsubsection{Variational Quantum Algorithms for Reference State}

The VQE is employed to approximate the ground state of the qubit Hamiltonian. VQE leverages the variational principle, minimizing the expectation value \(\langle \hat{H}_{\text{qubit}} \rangle\) over a parametrized quantum state \(|\psi(\boldsymbol{\theta})\rangle_{\text{QC}}\), prepared on a quantum circuit~\cite{Peruzzo2014}.
The ansatz state is generated by applying a unitary operator \(\hat{U}(\boldsymbol{\theta})\) to an initial reference state \(|\Phi_0\rangle\), typically the Hartree-Fock state~\cite{Peruzzo2014}:
\begin{equation}
|\psi(\boldsymbol{\theta})\rangle_{\text{QC}} = \hat{U}(\boldsymbol{\theta}) |\Phi_0\rangle.
\label{eq:ansatz_state}
\end{equation}
Common Ansatzes include the unitary coupled-cluster (UCC) form, such as UCCSD, which approximates the exponential cluster operator~\cite{Peruzzo2014}:
\begin{equation}
\hat{U}(\boldsymbol{\theta}) = \exp\left( \sum_i \theta_i \hat{\tau}_i - \theta_i^* \hat{\tau}_i^\dagger \right),
\label{eq:ucc_ansatz}
\end{equation}
where \(\hat{\tau}_i\) are excitation operators mapped to qubits.
The energy expectation value is
\begin{equation}
E(\boldsymbol{\theta}) = \langle \psi(\boldsymbol{\theta}) | \hat{H}_{\text{qubit}} | \psi(\boldsymbol{\theta}) \rangle_{\text{QC}} = \sum_k c_k \langle \hat{P}_k \rangle,
\label{eq:vqe_energy}
\end{equation}
evaluated by measuring the Pauli strings on the quantum device. Classical\added{-computing}\comment{R3 Q3} optimization algorithms, such as gradient descent or BFGS, minimize \(E(\boldsymbol{\theta})\) to find the optimal parameters \(\boldsymbol{\theta}^*\)~\cite{Tilly2022}.
Building upon the CAS and the VQE framework, the derived method provides a powerful hybrid approach for NISQ simulation. Here, VQE is applied to variationally solve the eigenvalue problem within the active space, serving as a quantum analogy of classical\added{-computing}\comment{R3 Q3} CASCI~\cite{Asthana2024}. classical\added{-computing}\comment{R3 Q3} CASCI expresses the wave function as a full linear combination of Slater determinants in the active space, and solve the eigen states via direct diagonalization. However, on the classical computer, the active space dimension grows factorially with the number of active orbitals and electrons, making classical\added{-computing}\comment{R3 Q3} CASCI intractable for large spaces. The CAS-VQE method make this tractable by exploiting quantum computers' ability to encode spin orbitals\cite{jpcacasvqe},  where VQE ansatz spans the Hilbert space of active orbitals and gives compact parametrization of correlations~\cite{Mizukami2020}. Optimization of the CAS-VQE follows standard VQE but employs the active space-restricted Hamiltonian, simulating only relevant electronic degrees of freedom. This enables accurate ground-state energies in static-correlation regimes while ensuring computational tractability.

Beyond the canonical single-reference VQE in selected active spaces, multi-reference feature can be incorporated to enhance the ground-state search in systems exhibiting strong static correlation. One such extension is the SSVQE, which expands the variational search into a larger subspace spanned by multiple orthogonal reference states~\cite{nakanishi2019subspace} in the selected active spaces.

In SSVQE, a set of $k$ mutually orthogonal initial reference states $\{|\phi_j\rangle\}_{j=1}^k$ is selected, often including the Hartree-Fock state and additional configurations to introduce multi-reference feature. The same parametrized unitary operator $\hat{U}(\boldsymbol{\theta})$ is applied to each reference, producing trial states $|\psi_j(\boldsymbol{\theta})\rangle = \hat{U}(\boldsymbol{\theta}) |\phi_j\rangle$. The variational cost function is defined as a weighted sum of the energy expectation values:
\begin{equation}
E(\boldsymbol{\theta}) = \sum_{j=1}^k w_j \langle \psi_j(\boldsymbol{\theta}) | \hat{H} | \psi_j(\boldsymbol{\theta}) \rangle,
\label{eq:ssvqe_cost}
\end{equation}
where the weights $w_j$ could chosen to be positive and decreasing ($w_1 > w_2 > \dots > w_k > 0$) to encourage the mapping of the trial states onto the lowest-energy eigenstates of the Hamiltonian. Classical\added{-computing}\comment{R3 Q3} optimization minimizes $E(\boldsymbol{\theta})$ to yield optimal parameters $\boldsymbol{\theta}^*$, effectively projecting the initial subspace onto the low-lying eigensubspace. The lowest-energy state among the optimized set, typically $|\psi_0(\boldsymbol{\theta}^*) \rangle$, serves as the variational ground state.

\subsubsection{Quantum Subspace Expansion Toolbox for Excited States}\label{section:qseee}

To access excited states in quantum\added{ computing} simulations, the QSE method projects the Hamiltonian into a subspace spanned by variationally prepared states and their excitations, followed by classical\added{-computing}\comment{R3 Q3} diagonalization of the subspace Hamiltonian to obtain excited states\cite{McClean2017}. This approach provides a physically-ensured way of calculating excited states with a given reference state, avoids the limitations of methods relying on parameter optimization.

In the Tamm-Dancoff Approximation for QSE (TDA-QSE)\cite{prreomqse}\label{tdaqse}, the subspace is constructed by applying low-rank excitation operators \( \hat{E}_\mu \)\deleted{to a reference state \( |\psi_k\rangle_{\text{QC}} \)} to a reference state \( |\psi_k\rangle_{\text{QC}} \), generating basis vectors, the excitations only include single excitations \( \hat{a}_p^\dagger \hat{a}_q \) and double excitations \( \hat{a}_p^\dagger \hat{a}_q^\dagger \hat{a}_r \hat{a}_s \):
\begin{equation}
|\phi_\mu\rangle = \hat{E}_\mu |\psi_k\rangle_{\text{QC}}.
\label{eq:qse_basis}
\end{equation}
\added{Note that the reference state \( |\psi_k\rangle_{\text{QC}} \) is prepared on quantum circuit by applying the optimal gate parameters (which are given by VQE) onto the Ansatz circuit (i.e. UCCSD).}\comment{R3 Q7}The projected Hamiltonian and overlap matrices are defined as:
\begin{equation}
H_{\mu\nu} = \langle \phi_\mu | \hat{H}_{\text{qubit}} | \phi_\nu \rangle, \quad S_{\mu\nu} = \langle \phi_\mu | \phi_\nu \rangle,
\label{eq:qse_matrices}
\end{equation}
\replaced{the Fermion operator $\hat{E}_\mu$ and $\hat{E}_\nu$ could be transformed and decomposed into $O(1)$ number of Pauli operators by Jordan Wigner transformation just how $\hat{H}_{\text{AS}}$ transformed into $\hat{H}_{\text{qubit}}$ in eq(\ref{eq:jw_transform}).}{where matrix elements are evaluated via quantum measurements}.\comment{R3 Q7} Excited-state energies and eigenvectors are obtained by solving the generalized eigenvalue problem:
\begin{equation}
\mathbf{H} \mathbf{c} = E \mathbf{S} \mathbf{c}.
\label{eq:qse_eig}
\end{equation}
The number of matrix elements scales as \( \mathcal{O}(m^8) \), where \( m \) is the number of selected orbitals. This results in an overall operator complexity of \( \mathcal{O}(m^4) \).

Beyond TDA-QSE, the general QSE framework supports a versatile set of fermionic operators in second quantization, enabling flexible subspace construction for systems with specific electron structures, such as those near conical intersections or avoided crossings. Each operator has an inverse (its Hermitian adjoint, e.g., de-excitations for excitations), which can enhance the description of physical properties\cite{jctc1}. Below, we list extensible operator classes, their second quantization forms, and their computational complexities, assuming a spin-orbital basis with \( m \) active spatial orbitals. Summations (e.g., \( p > q \)) ensure proper antisymmetrization, and spin-adapted forms\cite{gandon2024quantumcomputingspinadaptedrepresentations} are used where applicable to target singlet states (\( S^2 = 0 \)).

\begin{itemize}
    \item \textbf{Higher-Order Excitations}: These extend single and double excitations to triples, quadruples, and beyond, capturing higher-order electron correlations critical for multi-reference systems. For a triple excitation (from occupied orbitals \( i, j, k \) to virtual orbitals \( a, b, c \)):
    \begin{equation}
          \hat{E}_{ijk}^{abc} = \hat{a}_a^\dagger \hat{a}_b^\dagger \hat{a}_c^\dagger \hat{a}_k \hat{a}_j \hat{a}_i.  
    \end{equation}
    The inverse is \( \hat{E}_{abc}^{ijk} = \hat{a}_i^\dagger \hat{a}_j^\dagger \hat{a}_k^\dagger \hat{a}_c \hat{a}_b \hat{a}_a \). The number of triple excitation operators scales as \( \mathcal{O}(m^6) \), quadruples as \( \mathcal{O}(m^8) \), and general \( k \)-th order excitations as \( \mathcal{O}(m^{2k}) \), with de-excitations sharing the same complexity. Spin-adapted forms, constructed analogously to single and double excitations, commute with \( \hat{S}^2 \), reducing the constant factor in the subspace dimension.

    \item \textbf{Spin-Flip Operators}: These alter spin multiplicity by flipping electron spins while conserving orbital occupancy, useful for describing open-shell configurations. A single spin-flip (e.g., \( \alpha \) to \( \beta \) in orbitals \( p, q \)):
    \begin{equation}
          \hat{SF}_{p\beta,q\alpha} = \hat{a}_{p\beta}^\dagger \hat{a}_{q\alpha}.
    \end{equation}

    The inverse is \( \hat{a}_{q\alpha}^\dagger \hat{a}_{p\beta} \). Double spin-flips follow similarly (e.g., \( \hat{a}_{p\beta}^\dagger \hat{a}_{q\beta}^\dagger \hat{a}_{s\alpha} \hat{a}_{r\alpha} \)). Single spin-flips scale as \( \mathcal{O}(m^2) \), double spin-flips as \( \mathcal{O}(m^4) \). These operators do not naturally restrict to singlet states, as they couple different spin sectors, increasing the subspace dimension unless filtered by \( \hat{S}^2 \) commutation.

    \item \textbf{Spin-Mixing Operators}: These couple different spin sectors without flipping spins, enabling transitions between states of differing spin multiplicities while preserving total spin projection. A representative operator:
    \begin{equation}
            \hat{M}_{pq}^{rs} = \hat{a}_{p\alpha}^\dagger \hat{a}_{q\beta}^\dagger \hat{a}_{s\beta} \hat{a}_{r\alpha}.
    \end{equation}

    The inverse is \( \hat{a}_{r\alpha}^\dagger \hat{a}_{s\beta}^\dagger \hat{a}_{q\beta} \hat{a}_{p\alpha} \). These operators scale as \( \mathcal{O}(m^4) \) and, like spin-flip operators, do not inherently restrict to singlets, requiring \( \hat{S}^2 \) filtering to target singlet states.

    \item \textbf{Orbital Rotations}: These unitary transformations mix orbitals within the active space, optimizing orbital bases for response properties or strong correlation. The generator is anti-Hermitian:
    \begin{equation}
         \hat{\kappa} = \sum_{p>q} \kappa_{pq} \left( \hat{a}_{p\alpha}^\dagger \hat{a}_{q\alpha} - \hat{a}_{q\alpha}^\dagger \hat{a}_{p\alpha} + \hat{a}_{p\beta}^\dagger \hat{a}_{q\beta} - \hat{a}_{q\beta}^\dagger \hat{a}_{p\beta} \right).
    \end{equation}

    The inverse is \( -\hat{\kappa} \), as \( (\hat{\kappa})^\dagger = -\hat{\kappa} \). Orbital rotations scale as \( \mathcal{O}(m^2) \) and are naturally spin-adapted, commuting with \( \hat{S}^2 \), making them efficient for singlet subspaces.

    \item \textbf{Non-Diagonal Couplings}: These couple configurations across the orbital space without strict occupancy constraints, enhancing subspace flexibility. A general non-diagonal double coupling:
    \begin{equation}
            \hat{C}_{pq}^{rs} = \hat{a}_p^\dagger \hat{a}_q^\dagger \hat{a}_s \hat{a}_r.
    \end{equation}
    The inverse is \( \hat{a}_r^\dagger \hat{a}_s^\dagger \hat{a}_q \hat{a}_p \). These operators scale as \( \mathcal{O}(m^4) \) and can be spin-adapted (e.g., combining \( \alpha \)-\( \beta \) pairs symmetrically) to commute with \( \hat{S}^2 \), reducing the effective subspace size.

    \item \textbf{Electron-Electron Interaction Operators}: These directly incorporate two-body interaction terms from the molecular Hamiltonian, capturing dynamic and static electron correlations. Derived from the Hamiltonian’s two-electron integrals \( V_{pqrs} \). Including these operators enhances the subspace’s ability to describe correlation effects, but requires careful selection of significant integrals (e.g., \( |V_{pqrs}| > \epsilon \)) to manage computational cost. A representative operator is:
    \begin{equation}
            \hat{V}_{pqrs} = g_{pqsr}\hat{a}_p^\dagger \hat{a}_q^\dagger \hat{a}_s \hat{a}_r.
            \label{eeinterac}
    \end{equation}
    The inverse is \( \hat{a}_r^\dagger \hat{a}_s^\dagger \hat{a}_q \hat{a}_p \). The number of such operators scales as \( \mathcal{O}(m^4) \), matching double excitations. Spin-adapted forms, constructed similarly by ensuring commutation with \( \hat{S}^2 \). 
\end{itemize}

Implementation-level adaptions and optimizations could ameliorate the constant-factor overhead. When the number of selected expansion operators is large, the $S$ matrix may become ill-conditioned\cite{linlinqse}. We could adopt regularization methods to maintain numerical stability at the implementation level\cite{zhengzehua}. 

\added{Beyond its role in accessing excited states, QSE also serves as a powerful quantum error mitigation technique by projecting noisy or non-optimal reference states into a carefully constructed subspace that isolates and corrects errors inherent to near-term quantum hardware\cite{qem1,qem2,qem3}. In this context, QSE leverages excitation operators to expand the reference state, enabling the identification and suppression of noise-induced artifacts through the diagonalization of the projected Hamiltonian and overlap matrices, effectively restoring physical fidelity without requiring full error correction protocols. This application of QSE for error mitigation represents a distinct research branch, complementary to its excited-state computations, with ongoing developments focusing on adaptive operator selection and regularization to enhance robustness against quantum errors.}\comment{R3 Q18}

\subsubsection{Obtaining Nuclear Force}
To simulate the time evolution of the molecular system in the context of surface hopping dynamics, quality nuclear forces are essential. These forces are derived from the gradients of the PESs corresponding to different electronic states\cite{force1,force2}.

In our approach, the reference state $|\Psi_0\rangle$ is obtained using VQE or SSVQE. Excited states could be constructed using the result of QSE, where the excited state wavefunctions $|\Psi_k\rangle$ (for $k \geq 1$) could be written as the linear combinations of the reference state and applications of excitation operators $E_i$\cite{Takeshita_2020}:
\begin{equation}
   |\Psi_k\rangle = c_{k0} |\Psi_0\rangle + \sum_{i=1}^{M} c_{ki} E_i |\Psi_0\rangle, 
\end{equation}
where $\mathbf{c}_k = (c_{k0}, c_{k1}, \dots, c_{kM})^T$ is the $k$-th eigenvector from the QSE diagonalization, and $E_i$ are excitation operators. The coefficients $\mathbf{c}_k$ are obtained by solving the generalized eigenvalue problem in the QSE subspace. Note that this does not mean to explicitly prepare an excited-state wave function on quantum \replaced{computer}{register}, but through such representation, we could  finally construct observables and estimate corresponding expectations \added{as in \ref{expforce1}}\comment{R3 Q7}.

Nuclear forces are the negative gradients of the electronic energy with respect to nuclear coordinates $\mathbf{R}$:
\begin{equation}
    \mathbf{F}_k(\mathbf{R}) = -\nabla_{\mathbf{R}} E_k(\mathbf{R}),
\end{equation}

where $E_k(\mathbf{R}) = \langle \Psi_k | \hat{H}(\mathbf{R}) | \Psi_k \rangle$ is the energy of the $k$-th state, and $\hat{H}(\mathbf{R})$ is the molecular Hamiltonian.

Two primary methods exist for computing these gradients: finite difference method (FDM) and Hellman-Feynman method (HFM). 

The FDM provides a numerical gradient. It could approximate the gradient via central differences\cite{Pulay1983}
\begin{equation}
    \frac{\partial E_k}{\partial R_\alpha} \approx \frac{E_k(\mathbf{R} + \epsilon \mathbf{e}_\alpha) - E_k(\mathbf{R} - \epsilon \mathbf{e}_\alpha)}{2\epsilon},
    \label{eqfdm}
\end{equation}
where $\epsilon$ is the displacement step size, and $\mathbf{e}_\alpha$ is the unit vector along coordinate $\alpha$. 

The HFM provides analytical gradients. The Hellman-Feynman theorem states that for an exact or variationally optimized wavefunction $|\Psi\rangle$ of a Hamiltonian $\hat{H}(\lambda)$ depending on a parameter $\lambda$, the derivative of the energy $E(\lambda) = \langle \Psi | \hat{H} | \Psi \rangle$ (assuming normalization $\langle \Psi | \Psi \rangle = 1$) with respect to $\lambda$ is given by\cite{Levine2009,Lai2023}:
\begin{equation}
  \frac{dE}{d\lambda} = \left\langle \Psi \middle| \frac{\partial \hat{H}}{\partial \lambda} \middle| \Psi \right\rangle.  
  \label{eqhfm}
\end{equation}
This result follows from differentiating the energy expression:
\begin{equation}
    \frac{dE}{d\lambda} = \left\langle \frac{\partial \Psi}{\partial \lambda} \middle| \hat{H} \middle| \Psi \right\rangle + \left\langle \Psi \middle| \frac{\partial \hat{H}}{\partial \lambda} \middle| \Psi \right\rangle + \left\langle \Psi \middle| \hat{H} \middle| \frac{\partial \Psi}{\partial \lambda} \right\rangle.
\end{equation}

If $|\Psi\rangle$ is a normalized eigenstate of $\hat{H} $, the sum $\left\langle \frac{\partial \Psi}{\partial \lambda} \middle| \hat{H} \middle| \Psi \right\rangle + \left\langle \Psi \middle| \hat{H} \middle| \frac{\partial \Psi}{\partial \lambda} \right\rangle$ vanishes, as required by the eigenvalue equation $(\hat{H} - E) |\Psi\rangle = 0$  and the normalization condition  $\langle \Psi | \Psi \rangle = 1$ . Furthermore, upon adopting a gauge where the global phase of  $|\Psi(\lambda)\rangle$ is chosen such that $\langle \Psi | \frac{\partial \Psi}{\partial \lambda} \rangle = 0$, each term vanishes individually. For a variationally prepared wavefunction, if it is optimal with respect to its parameters, the response term will be zero by the variational principle, making the theorem applicable to states given by VQE or QSE\cite{Lai2023}.

In implementation, we would use finite basis set to construct wave function. When using atom-centered basis sets (e.g., Gaussian orbitals), the basis functions will depend implicitly on nuclear positions, and since $\lambda$ corresponds to nuclear coordinates $R_\alpha$, the force will be $-\langle \Psi | \partial \hat{H} / \partial R_\alpha | \Psi \rangle$. This would introduce additional contributions known as Pulay terms or Pulay forces. These arise because the derivative must account for the basis set's coordinate dependence\cite{Pathak2023}:
\begin{equation}
\frac{\partial \hat{H}}{\partial R_\alpha} = \frac{\partial \hat{H}}{\partial R_\alpha}\bigg|_{\chi} + \sum_{\mu\nu} \frac{\partial \chi_\mu}{\partial R_\alpha} \hat{h}_{\mu\nu} + \cdots,
\end{equation}
where $\chi$ denotes basis functions, and the Pulay correction includes terms from the overlap matrix derivatives and density matrix responses. In practice, for MO-based methods, the force operators incorporate these via the gradient of the core Hamiltonian and electron-repulsion integrals in the MO basis, augmented by Pulay contributions from the atomic orbital (AO) to MO transformation\cite{Pulay1983}.

The derivative Hamiltonian $\partial \hat{H}/\partial R_\alpha$ is thus computed in the MO basis, incorporating electron integrals transformed via the MO coefficients from a preceding Hartree-Fock calculation. Specifically, the force operators are derived as\cite{OBrien2019}:
\begin{equation}
    \hat{F}_\alpha = \sum_{pq} h_{pq}^\alpha a_p^\dagger a_q + \frac{1}{2} \sum_{pqrs} v_{pqrs}^\alpha a_p^\dagger a_r^\dagger a_s a_q,
\end{equation}
where $h_{pq}^\alpha$ and $v_{pqrs}^\alpha$ are the derivative core-Hamiltonian and electron-repulsion integrals, respectively, including Pulay terms for basis set dependence on nuclear positions. \added{In our implementation, we compute the energy gradients using the Hellmann-Feynman theorem by obtaining the derivative of one and two electron integrals (including Pulay terms) on classical computer.}\comment{R2 Q21}

Finally, on quantum computer, these Fermionic operators are mapped to qubit operators using the Jordan-Wigner transformation, and expectation values are evaluated via:
\begin{equation}
    \langle \psi_k | F_{\alpha} | \psi_k \rangle = \sum_{i,j} (c_j^k)^* c_i^k \langle \psi_{\mathrm{VQE}} | \hat{E}_j^\dagger A \hat{E}_i | \psi_{\mathrm{VQE}} \rangle
    \label{expforce1}
\end{equation}

\subsubsection{Quantum Computation for Molecular Hessian}
For surface hopping dynamics, second derivatives of the energy (the Hessian matrix) are crucial for computing vibrational frequencies, and ensuring initial conditions for the propagation\cite{Ibele2025, Barbatti2022, Suchan2018}. The Hessian element for coordinates $\alpha$ and $\beta$ is:
\begin{equation}
H_{\alpha\beta} = \frac{\partial^2 E_k}{\partial R_\alpha \partial R_\beta} = \frac{\partial F_{k,\beta}}{\partial R_\alpha}.
\end{equation}
Direct analytical computation of the Hessian on quantum \replaced{computer}{hardware} is challenging due to the need for higher-order responses. Instead, we could compute the Hessian via finite differences of the Hellmann-Feynman gradients:
\begin{equation}
    H_{\alpha\beta} \approx \frac{F_{k,\beta}(\mathbf{R} + \epsilon \mathbf{e}_\alpha) - F_{k,\beta}(\mathbf{R} - \epsilon \mathbf{e}_\alpha)}{2\epsilon}.
\end{equation}
Here, $\epsilon$ should be properly chosen to balance numerical stability and accuracy, minimizing the deviation propagation from the electronic structure solver while capturing curvature.

For each displacement, the molecular geometry is updated, and a new Hartree-Fock calculation provides updated MO coefficients. The calculation of electronic-structure observables are then repeated to obtain the ground-state energy and gradient at the perturbed geometry. The full Hessian is assembled as a $(3N \times 3N)$ matrix ($N:$ the number of atoms).

\subsection{Time evolution as a surface hopping dynamics}
\comment{R3 Q8}
\begin{figure}
    \centering
    \includegraphics[width=0.78\linewidth]{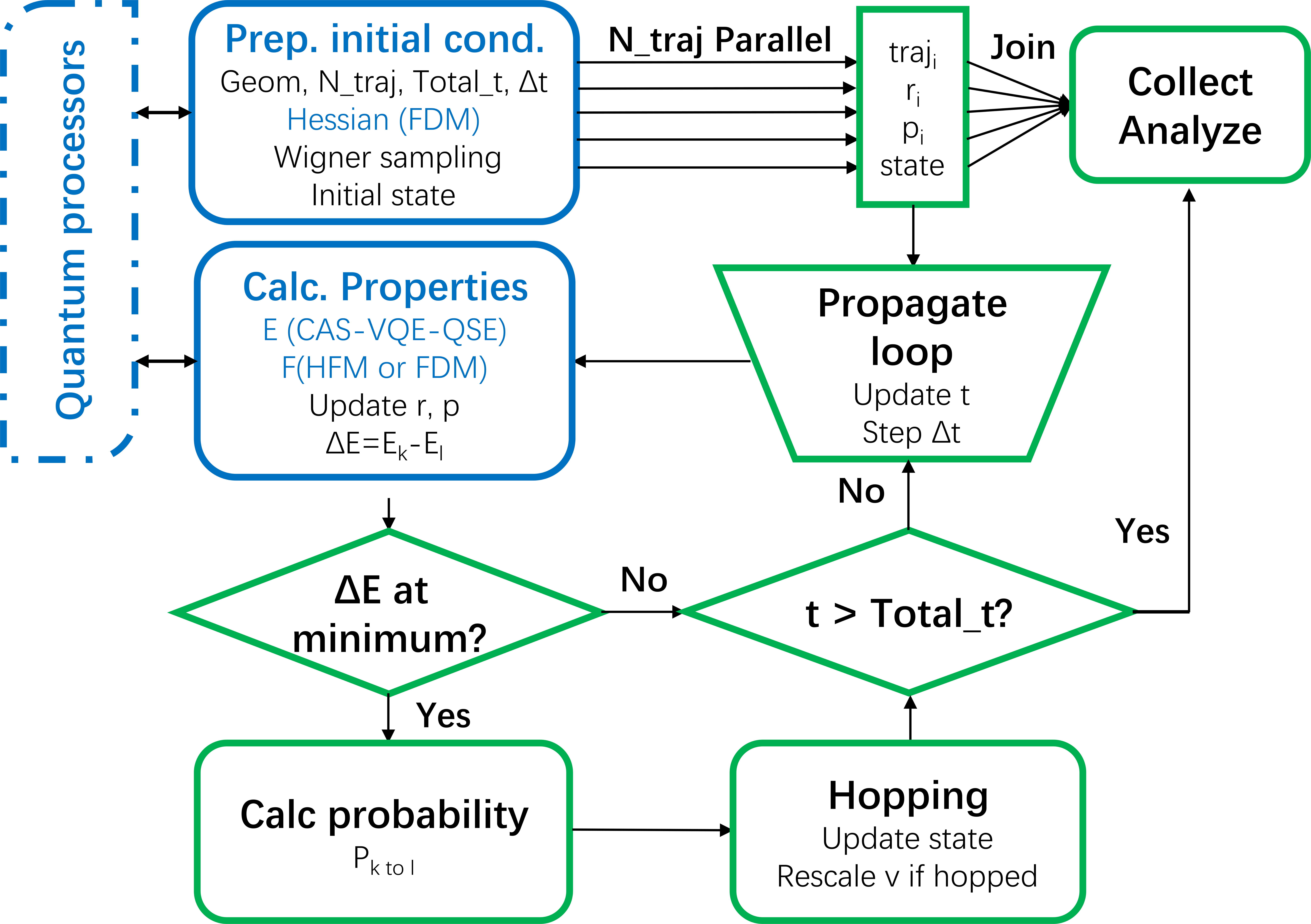}
    \caption{Work flow of parallel LZSH dynamics (represented by eq.(\ref{lzprobb})) that co-operates with on-the-fly quantum-computing electronic-structure property solver.}
\end{figure}
\subsubsection{Wigner Sampling and Landau-Zener Surface Hopping Dynamics}
To propagate the nuclear dynamics while accounting for non-adiabatic effects, we initialize an ensemble of classical trajectories (\added{in the classical physics sense, they were also computed on classical computer})\comment{R3 Q3} using Wigner sampling and employ a LZSH algorithm. This combination allows for the incorporation of initial quantum\added{-mechanical}\comment{R3 Q3} nuclear effects and efficient treatment of electronic state transitions without requiring explicit computation of NAC vectors, which is particularly efficient.

Wigner sampling provides a phase-space distribution that approximates the quantum mechanical density for the initial vibrational state, typically the ground state at zero temperature \cite{McQuarrie2000}. For a molecule treated as a set of harmonic oscillators derived from the Hessian matrix, the Wigner distribution in normal coordinates $\mathbf{Q}$ and conjugate momenta $\mathbf{P}$ is given by
\begin{equation}
    \rho_W(\mathbf{Q}, \mathbf{P}) = \prod_{l=1}^{3N-6} \frac{\alpha_l}{\pi \hbar} \exp\left( -\frac{2}{\hbar \omega_l} \left( \frac{P_l^2}{2} + \frac{1}{2} \omega_l^2 Q_l^2 \right) \right),
\end{equation}
where $\omega_l$ is the frequency of the $l$-th normal mode, and $\alpha_l = \tanh(\hbar \omega_l / 2 k_B T)$ approaches 1 at $T=0$ K (the regime considered here) \cite{Kammerer2002}. For the ground state, each mode's position $Q_l$ and momentum $P_l$ (scaled by the reduced mass) are independently sampled from Gaussian distributions:
\begin{equation}
   Q_l \sim \mathcal{N}\left(0, \frac{\hbar}{2 \omega_l}\right), \quad P_l \sim \mathcal{N}\left(0, \frac{\hbar \omega_l}{2}\right). 
\end{equation}

On the classical computer, these normal-mode samples are transformed back to Cartesian coordinates and velocities using the eigenvectors from the Hessian diagonalization, ensuring the initial ensemble captures zero-point energy and quantum\added{-mechanical}\comment{R3 Q3} delocalization effects \cite{McQuarrie2000}. 
The sampled trajectories are then propagated on the adiabatic PESs computed via VQE and QSE \cite{Tilly2022,Bian2021}. Non-adiabatic transitions are handled via the LZSH algorithm, a computationally efficient variant of Tully's FSSH that approximates hopping probabilities using the Landau-Zener formula without needing time-dependent electronic coefficients or NAC \cite{Tully1990}. In LZSH, at each time step $\Delta t$, for the current active state $k$ and each other state $l \neq k$, the energy gap $\Delta_{kl} = |E_k - E_l|$ is monitored. A hop is considered only if the gap reaches a local minimum (i.e., $\Delta_{kl}(t) > \Delta_{kl}(t - \Delta t)$ and $\Delta_{kl}(t) < \Delta_{kl}(t + \Delta t)$ in a retrospective check), indicating passage through an avoided crossing \cite{Barbatti2011}. The hopping probability from $k$ to $l$ is then given by the Landau-Zener formula for the transition probability \cite{Hagedorn1991,Wang2014}:

\begin{equation}
    P_{k \to l} = \exp\left( -\frac{\pi}{2\hbar} \sqrt{\frac{\Delta_{kl,min}^3}{|\ddot{\Delta}_{kl,min}|}} \right),\label{lzprobb}
\end{equation}\comment{R2 Q10,ik to kl} 
In the implementation, if a random number $\xi \in [0,1)$ is less than $P_{k \to l}$, a hop occurs, and the velocity is rescaled along the force difference direction to conserve energy; otherwise, the trajectory continues on the current surface \cite{Sisourat2015}. 

\added{A challenge of the LZSH NAMD is its sensitivity to the choice of nuclear time step, and could be less reliable for systems involving more than two electronic states due to oversimplified multi-state interactions.} \comment{R2 Q17}Another critical challenge in LZSH NAMD is the presence of discontinuities in the PESs, which can manifest as artificial local minima in the energy gap when dynamics reach the far end of the dissociation region where PESs are closely spaced. These artifacts, arising from numerical instabilities in electronic structure solvers (despite high accuracy), lead to erroneous transition probabilities and non-physical dynamics. Hard-coded filtering risks losing valuable information, so, inspired by \citet{Jira2025}, we implement a curvature-induced transition protection algorithm tailored for quantum\added{-computing}\comment{R3 Q3} electronic solvers. This approach suppresses spurious transitions from PES fractures over small displacement intervals, relying solely on energy gap information already available in the LZSH procedure (thus requiring no additional electronic-structure calculations) and efficiently refines the dynamics for more physically consistent system evolution.

Specifically, the algorithm computes a coefficient $\alpha$ that measures the relative change in the second derivative (curvature) of the energy gap $\Delta_{kl}$ between the step immediately preceding the detected minimum and the minimum itself:
\begin{equation}
   \alpha = \left|\frac{\ddot{\Delta}_{kl, \text{prev}} - \ddot{\Delta}_{kl, \text{min}}}{ \ddot{\Delta}_{kl, \text{min}}}\right| ,
\label{curvaturecorrection}
\end{equation}
where \(\ddot{\Delta}_{kl, \text{prev}}\) and \(\ddot{\Delta}_{kl, \text{min}}\) are the curvatures at those respective steps. The purpose of $\alpha$ is to detect whether the minimum is likely an unphysical artifact (e.g., from a discontinuity-induced fracture) rather than a genuine physical feature: a large $\alpha$ indicates an abrupt, suspicious change in curvature. If \(\alpha > c_{block}\), the hop is blocked as the minimum is deemed discontinuity-induced (i.e., a trivial crossing); if \(c_{alert} \leq \alpha \leq c_{block}\), a warning is issued (potentially flagging a sharp but physical conical intersection, akin to a nontrivial crossing); otherwise, the hop proceeds normally. The thresholds \(c_{alert}\) and \(c_{block}\) are empirical values, summarized as 0.3 and 1.3\cite{Jira2025}. \added{This safeguard promotes smooth curvature changes to eliminate spurious crossings arising from numerical artifacts, while preserving physically meaningful ones, thereby improving stability without resorting to hard-coded filters or additional quantum evaluations.}\comment{R2 Q12}

\subsection{Implementations}
Surface-hopping NAMD simulations require repeated execution of electronic structure solvers. In \deleted{classical simulations of }subspace quantum electronic structure solvers, QSE incurs a significant cost, requiring the estimation of \( \mathcal{O}(m^8) \) matrix elements. \deleted{On the engineering level}, inspired from the serial prototype\cite{Chemqulacs2025}, noting that the computation of QSE matrix elements in Eqs.~\eqref{eq:qse_matrices} can be parallelized. Each expectation value \(\langle \hat{P}_k \rangle\) for Pauli strings in decomposed operators can be estimated independently. We leverage this by parallel computations across multiple processors for classical\added{-computing} \added{quantum algorithm} simulation or real-device quantum computing\comment{R3 Q3}. On the \replaced{theoretical}{physics} level, we exploit the Hermitian feature of QSE matrices to halve the matrix estimation overhead, which is a general strategy. Another optimization applies to systems where dynamics simulations involve only singlet excited-state energies: select operators that commutes with \(\hat{S}^2\), we achieve a fourfold reduction in operator count. Furthermore, for specific target states in particular systems, operators with negligible contributions could also be eliminated based on their respective physical nature, enabling problem-specific operator savings.

The surface hopping framework requires a sufficiently large ensemble of trajectories to mitigate statistical noise and accurately capture the underlying physical behavior. Given the resource-intensive nature of quantum alogirthm simulators, we leverage the inherent independence of trajectories within the ensemble to enhance computational efficiency on the engineering level.  Each trajectory, initialized from the Wigner distribution by calling \replaced{the program immigrated from Newton-X}{Newton-X pragram}\added{-2.4-B06}\comment{R1 Q4}\cite{newtonx}, evolves autonomously under the surface hopping dynamics, allowing for parallelization across multiple processors or computational nodes. In implementation, the initial phase-space points ${\mathbf{Q}_l, \mathbf{P}_l}$ \comment{R2 Q10, i to l }are assigned to parallel workers. Each worker propagates its assigned subset of trajectories forward in time using the classical\added{-computing}\comment{R3 Q3} integrator on the active surface, interspersed with hop evaluations at each step. The quantum computations for energies $E_k$ and forces $F_k$ are invoked on-demand for each trajectory's current geometry.\replaced{ On the other hand}{On the physical level}, the curvature-induced hopping correction could maintain computationally efficiency. The parameter $\alpha$ in eq.(\ref{curvaturecorrection}) can be obtained using a \added{backward difference}\comment{R2 Q11} of the energy-gap curvatures between the current and previous time steps, eliminating the need for additional wavefunction calculations beyond those already performed in the standard LZSH propagation. \added{The curvature at each time step is computed via central difference of the energy gaps at neighboring steps:}\begin{equation}
    \ddot{\Delta}_{kl, t}= \frac{\Delta_{t+1} + \Delta_{t-1} - 2 \Delta_{t}}{(\tau)^2}
\end{equation}
\added{where $\tau$ is the time step size. Note that the energy gap at the trial step $t+1$ is already calculated in the standard LZSH algorithm to locate the local minimum that triggers surface hopping according to Eq.~(\ref{lzprobb}. By estimating $\alpha$ using a backward difference of curvatures (relying only on information up to step $t+1$), we avoid any additional propagation to a hypothetical $t+2$ step that would otherwise be required for a forward or central difference.}\comment{R2 Q11}

The simulator programs for subspace quantum algorithms are implemented using the MindSpore Quantum package~\cite{xu2024mindsporequantumuserfriendlyhighperformance}, Quri-parts\cite{quri-parts}, and Qulacs\cite{qulacss}.
To construct our \replaced{hybrid quantum-classical (both in physical sense and computational sense)}{quantum-classical hybrid}\comment{R3 Q3} program, we referred to a canonical implementation of classical AIMD framework in MLatom\cite{MLatom1,MLatom2,MLatom3,MLatomProg} as an initial baseline.

We use PySCF \added{2.8.0}\comment{R1 Q4} \cite{pyscf} do the classical\added{-computing}\comment{R3 Q3} reference calculation at the same level corresponding to the quantum\added{-computing}\comment{R3 Q3} solver as the exact solution. All electronic-structure property calculations employ the STO-3G basis set. We pick UCCSD Ansatz\cite{Anand_2022} \added{for C$_2$H$_4$ use case} and k-UpUCCGSD Ansatz\cite{kup} \added{for H$_3^+$ use case}\comment{R2Q13} respectively. For VQE parameter optimizer, we use LBFGS\cite{liu1989limited,zha2021impactsoptimizationalgorithmbasis}. \added{In the LZSH program, we pick time step be 0.2 fs. On the Wigner sampling for the initial condition preparation, we assume a simple $0$k temperature and $\delta$ impulse for excitation, without the filtering by excitation window. During the NAMD propagation, velocities are rescaled uniformly along their current direction upon a successful hop to compensate for the energy difference between the target and initial electronic states, with the kinetic energy adjusted by the negative of this gap to maintain total energy conservation. Frustrated hop will leave the trajectory at the current electronic state. At the step of curvature-induced hopping correction, given that the instability only occurs at the final stage of dissociation where PESs are very close (see \ref{popblock}), we set \(c_{alert} = 0.3\) and \(c_{block} = 0.9\). }\comment{R2 Q14,15,18}

\clearpage
\section{Results }

In the first subsection, we focus on the 3-orbital-2-electron (CAS(3,2)) space with the charged H$_3^+$ ion. For this system, we compare PESs along dissociation geometries using the TDA-QSE (\added{as introduced for eq.(\ref{eq:qse_basis})}\comment{R3 Q10}, we denote as "QSE" in the following discussions for simplicity) and operator extended QSE methods (\added{e.g. electronic-electronic interaction in eq.(\ref{eeinterac}) or other operators introduced}\comment{R3 Q10}, we denote as "QSE*"), highlighting the systematic accuracy improvement of the QSE* approach. We then evaluate two gradient computation methods (FDM and HFM) under QSE and QSE*, including a comparison of different FDM step lengths. Additionally, we present LZSH-NAMD simulation results using QSE and QSE* as electronic structure solvers, augumented with the curvature-induced hopping correction.

In the second subsection, we would demonstrate adapatbility by focusing on a larger molecule C$_2$H$_4$. Using hybrid \replaced{subspace quantum-computing electronic structure solvers}{subspace-based quantum solvers}\comment{R3 Q3}, we capture key chemical properties within a selected 2-orbital-2-electron active space (CAS(2,2)): the conical intersection during the 'pyramidalization' process of C$_2$H$_4$. We validate high-accuracy quantum-computed PESs along the model path on quantum simulator, followed by nuclear forces computed via FDM. Finally, we validate our method along a NAMD trajectory.

We compute all the PESs and forces of both H$_3^+$ and C$_2$H$_4$ without geometry symmetry assumption, which validates the practicality of the electronic structure solvers in NAMD. \added{We selected small basis set and active spaces to balance computational resource, which necessitate numerous repeated quantum algorithm executions on classical-computing simulators. Larger basis set or active spaces would increase qubit requirements, substantially prolonging VQE and QSE times, escalating overall computation demands despite our parallel optimizations.}\comment{R2Q16} \added{We opt Hartree as the energy unit which is related to the electronvolt (eV) by the conversion factor of 1 Hartree $\approx$ 27.211 eV.}\comment{R3Q20}

In rest subsections, to further evaluate the near-term practical potential, we calculated these observables with noisy quantum \added{algorithm} simulator \added{on classical computer}\comment{R3Q3}. As a complement to the long-term potential, we validated some PES calculations for triplet states of H$_3^+$ and CH$_2$O, preliminarily exploring quantum simulations of open-shell systems. Finally, we validate the speedup of our two-level parallelization framework by comparing to the serial versions.

\subsection{Use case I: H$_3^+$ Results in CAS(3,2)}
\begin{figure}[h!]
  \centering
  \begin{subfigure}[b]{0.46\textwidth}
    \centering
    \includegraphics[width=\textwidth]{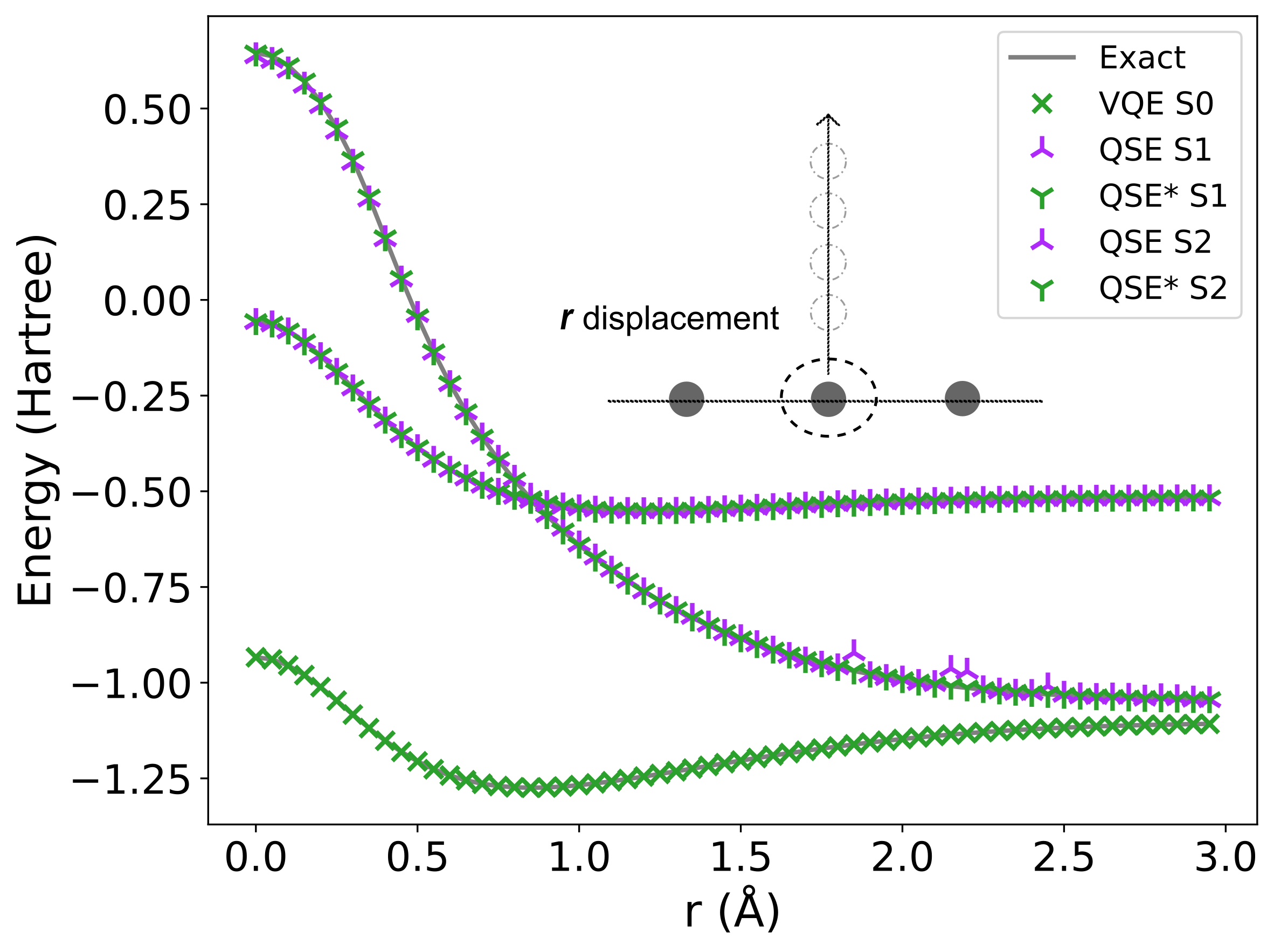}
\caption{}
   
  \end{subfigure}
  \hfill
  \begin{subfigure}[b]{0.46\textwidth}
    \centering
    \includegraphics[width=\textwidth]{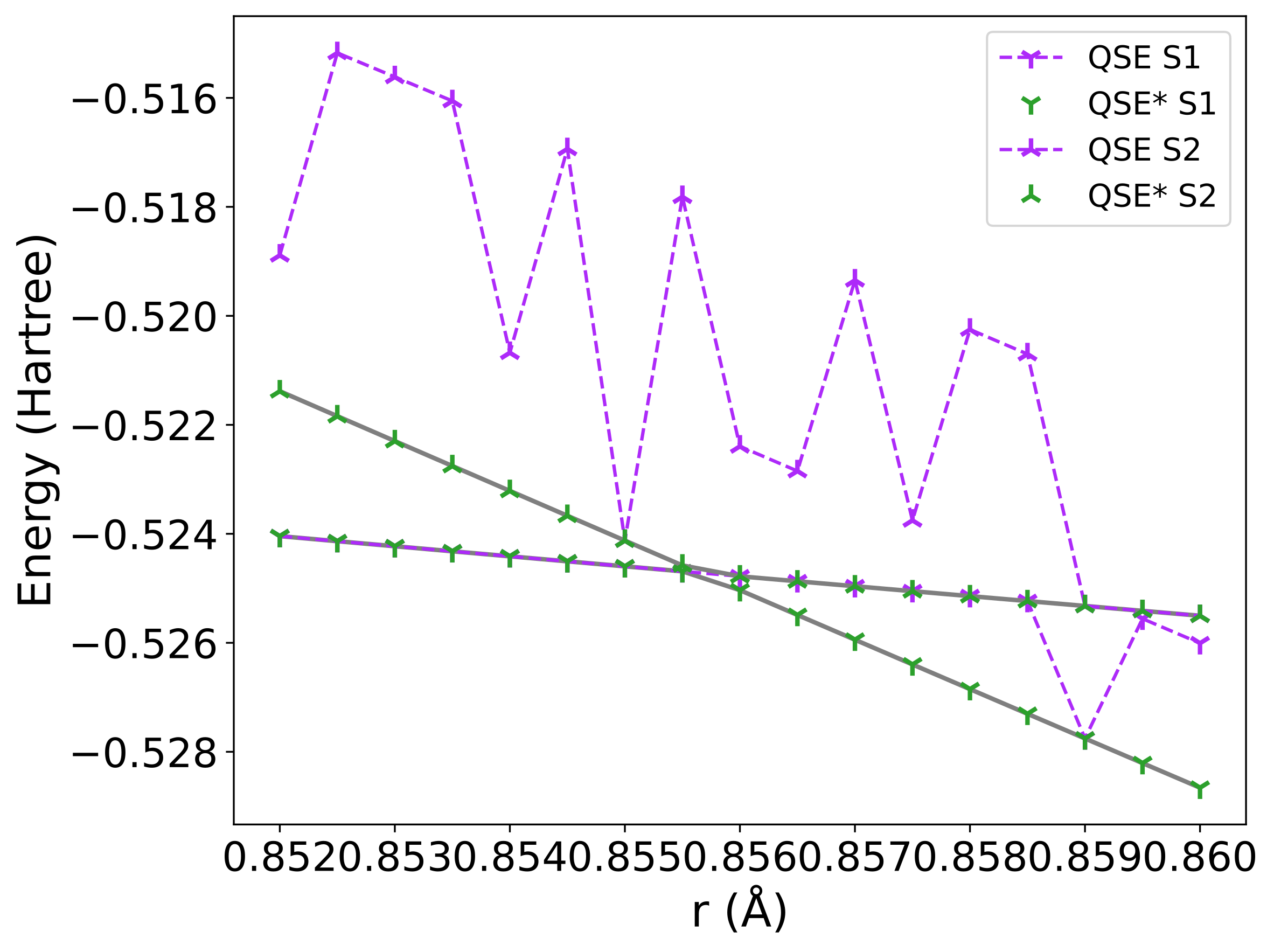}
\caption{}

  \end{subfigure}

  \caption{PES result along demonstrative disassociation geometries of H$_3^+$ in CAS(3,2). Gray lines represent the reference results, purple markers denote the singlet-adapted QSE results, green markers denote the singlet-adapted QSE* reuslts (augmented with the electron-electron interaction operators that preserve singlet spin multiplicity, as introduced in eq.(\ref{eeinterac})). (a) PES results compares two methods, with an interval of $0.05$ Å between adjacent points. (b) PES results around the intersection region comparing two methods, with an interval of $0.0005$ Å.}

\label{fig:h3p_pes}
\end{figure}

\begin{table}[h]
\centering
\caption{Comparison of PES errors with different interval length of different solvers for H$_3^+$ in CAS(3,2). }
\begin{tabular}{lccccc}
\toprule
$\Delta$E (Hartree) & VQE S$_0$ & QSE* S$_1$ & QSE*  S$_2$ & QSE S$_1$ & QSE S$_2$ \\
\midrule
\multicolumn{6}{c}{$0.05$ Å interval} \\
RMSE & \num{9.59e-14} & \num{2.02e-13} & \num{1.49e-14} & \num{9.23e-03} & \num{5.40e-03} \\
Max Error & \num{4.49e-13} & \num{1.34e-12} & \num{8.56e-14} & \num{4.88e-02} & \num{3.14e-02} \\
MAE & \num{3.87e-14} & \num{6.62e-14} & \num{5.73e-15} & \num{3.35e-03} & \num{2.02e-03} \\
\midrule
\multicolumn{6}{c}{$0.0005$ Å interval} \\
RMSE & \num{6.97e-14} & \num{1.48e-14} & \num{1.50e-14} & \num{1.20e-03} & \num{4.37e-03} \\
Max Error & \num{1.39e-13} & \num{3.31e-14} & \num{4.56e-14} & \num{2.65e-03} & \num{6.77e-03} \\
MAE & \num{5.12e-14} & \num{8.87e-15} & \num{6.29e-15} & \num{7.22e-04} & \num{3.49e-03} \\
\bottomrule
\end{tabular}
\label{tabpesh3ppes}
\end{table}

The PESs for the H$_3^+$ cation in its CAS(3,2) space were computed along a dissociation coordinate, where one hydrogen atom is displaced from the equilateral triangular equilibrium geometry. This path encompassing regions of conical intersection between the $S_1$ and $S_2$ excited states near $r \approx 0.85 $Å. The singlet adapted VQE (\added{single reference version, as introduced eq.(\ref{eq:vqe_energy})}\comment{R3 Q10}, with k-UpUCCGSD Ansatz) was employed for the $S_0$ state, while the singlet adaption, in both QSE and QSE* variants, was utilized for the $S_1$ and $S_2$ states. The QSE* approach incorporating additional electron-electron interaction operators, thus provides a more accurate description of electron correlation energy for ion system.

Figure~\ref{fig:h3p_pes}(a) illustrates the PESs over an extended dissociation range ($0-3 $Å) with an interval of $0.05$ Å between adjacent points, comparing the computed energies against exact FCI references \added{(note that for H$_3^+$, CAS(3,2) already takes all the electrons and orbitals, so here in this case, UCCSD equals to FCI)}\comment{R3 Q11}. The VQE $S_0$ curve closely tracks the exact ground-state PES, exhibiting a deep potential well at equilibrium bond length, followed by a smooth rise to the dissociation limit. For the excited states, the QSE method yields noticeable deviations, which fails to accurately capture the critical behavior, resulting in oscillations and energy offsets. In contrast, the QSE* method demonstrates good fidelity, with both $S_1$ and $S_2$ curves overlaying the exact references across the entire coordinate, including the flat dissociation plateau beyond $r \approx 2.0 $Å.

A magnified view around the conical intersection region ($0.852-0.86$Å) is provided in Figure~\ref{fig:h3p_pes}(b), highlighting the PES accuracy even with an interval of 0.0005 Å between adjacent points. The QSE* results reproducing the conical intersection with high precision, whereas the QSE introduces erratic fluctuations indicative of subspace incompleteness due to inadequate selection of operators.

Quantitative energy errors, summarized in Table~\ref{tabpesh3ppes}, further underscore these observations. For the PES with interval of $0.5$ Å, VQE achieves root-mean-square error (RMSE) and mean absolute error (MAE) values on the order of $10^{-14} \text{ Hartree}$, affirming its robustness for ground-state simulations. The extended method attains comparable sub-microhartree accuracy for $S_1$ (RMSE = $2.02 \times 10^{-13} \text{ Hartree}$) and $S_2$ (RMSE = $1.49 \times 10^{-14} \text{ Hartree}$), with maximum errors below $10^{-11} \text{ Hartree}$-orders of magnitude better than the QSE, which incurs RMSEs of $9.23 \times 10^{-3} \text{ Hartree}$ and $5.40 \times 10^{-3} \text{ Hartree}$ for $S_1$ and $S_2$, respectively. Similar trends persist in the regime with interval of $0.0005$ Å, where QSE* errors remain at the $10^{-14} \text{ Hartree}$ level, while QSE errors escalate to millihartree scale (e.g., RMSE = $4.37 \times 10^{-3} \text{ Hartree}$ for $S_2$), reflecting its numerical instability near the intersection.

\begin{figure}[h]
    \centering
    \begin{subfigure}{0.45\textwidth}
        \includegraphics[width=\textwidth]{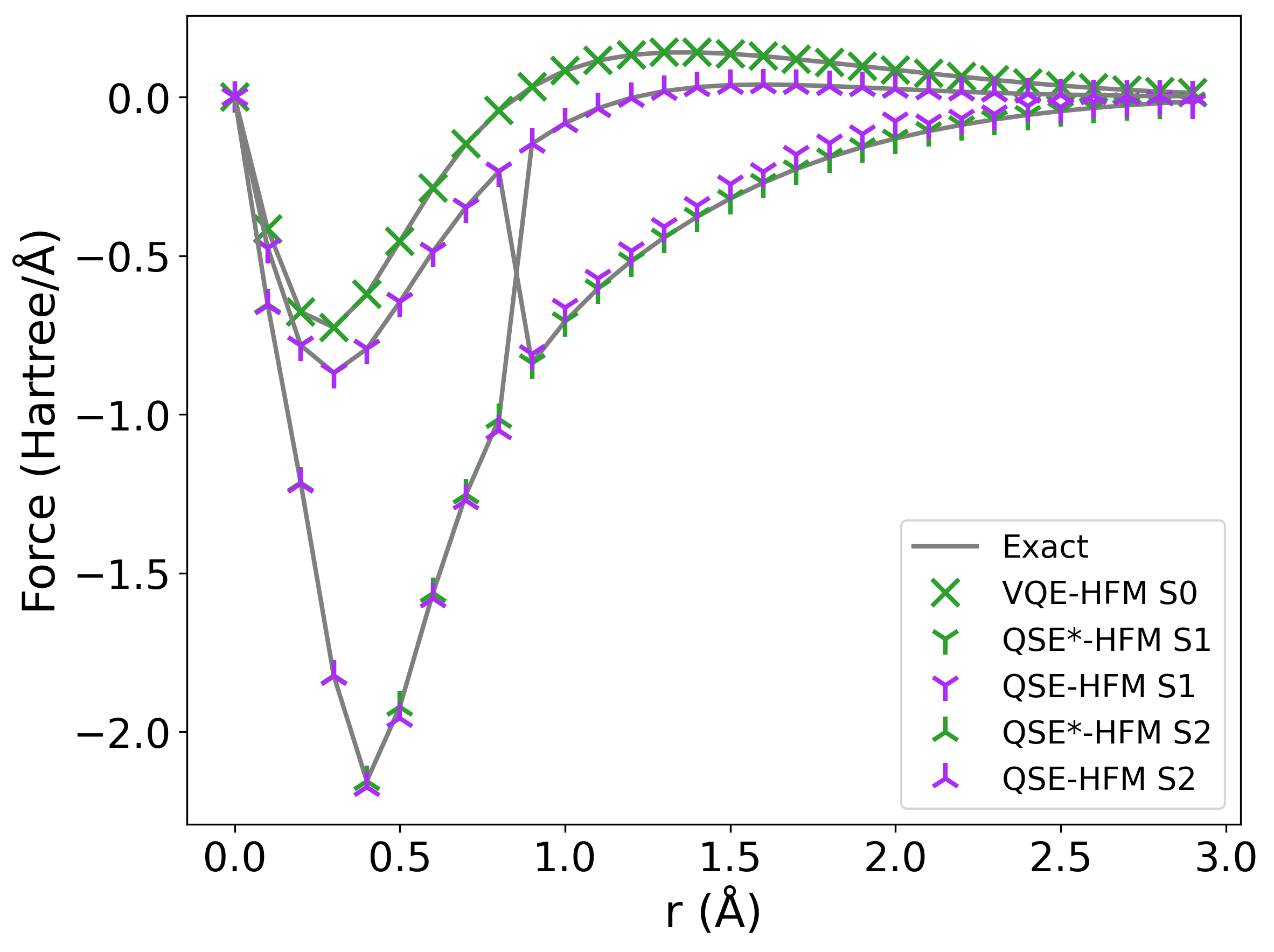}
        \caption{}
    \end{subfigure}
    \hfill
    \begin{subfigure}{0.45\textwidth}
        \includegraphics[width=\textwidth]{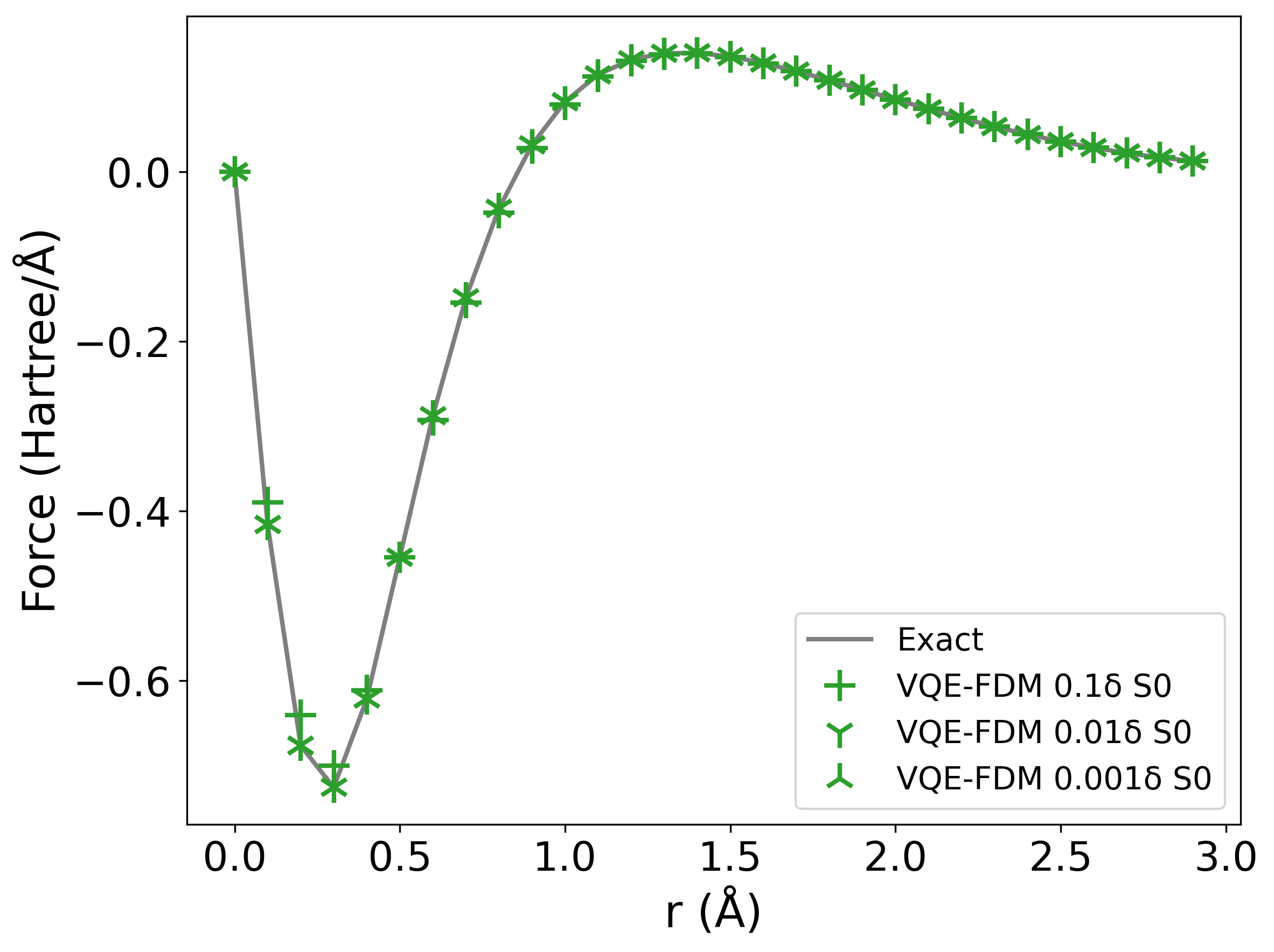}
        \caption{}
    \end{subfigure}

    \vspace{0.5cm} 

    \begin{subfigure}{0.45\textwidth}
        \includegraphics[width=\textwidth]{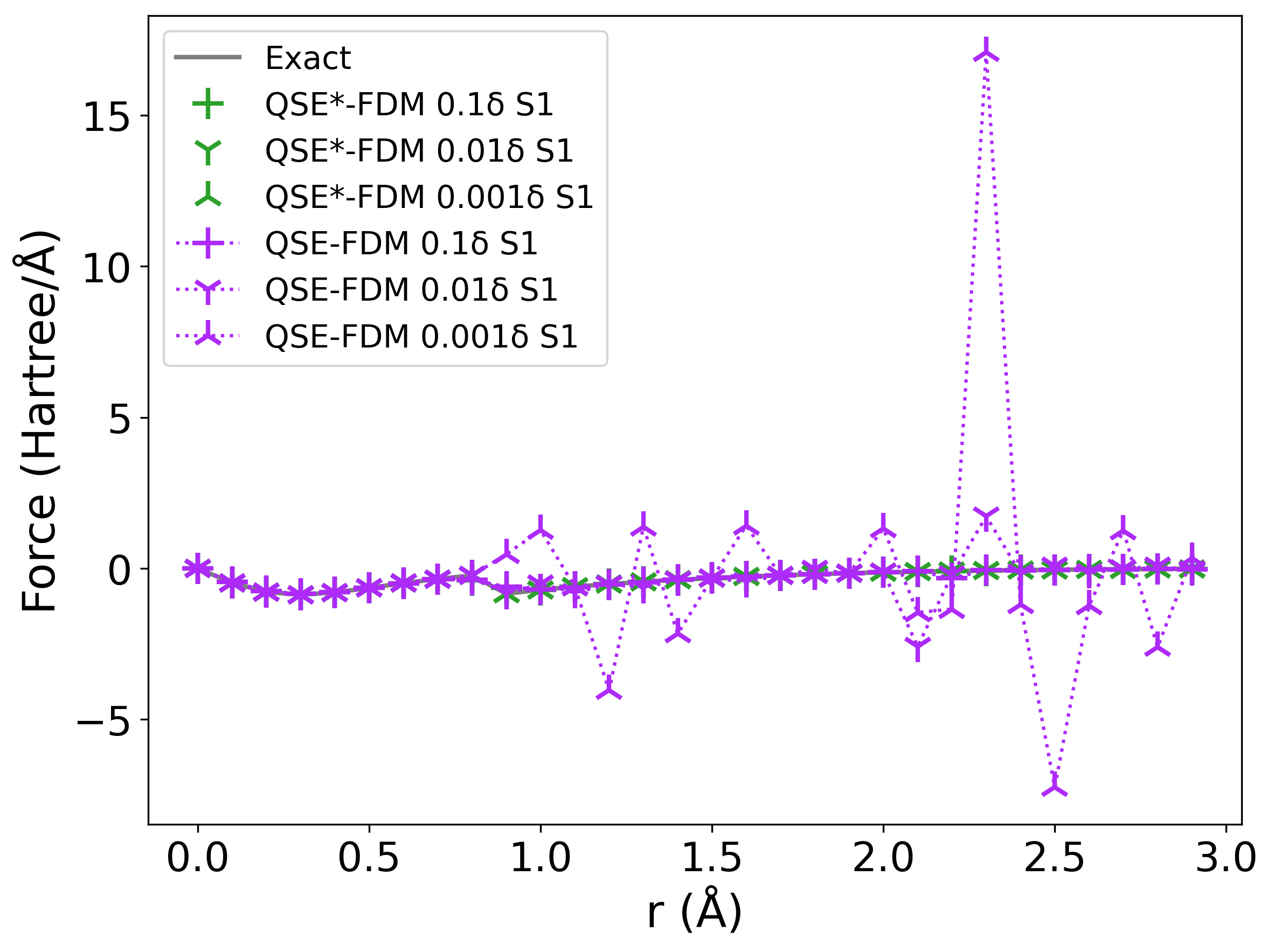}
        \caption{}
    \end{subfigure}
    \hfill
    \begin{subfigure}{0.45\textwidth}
        \includegraphics[width=\textwidth]{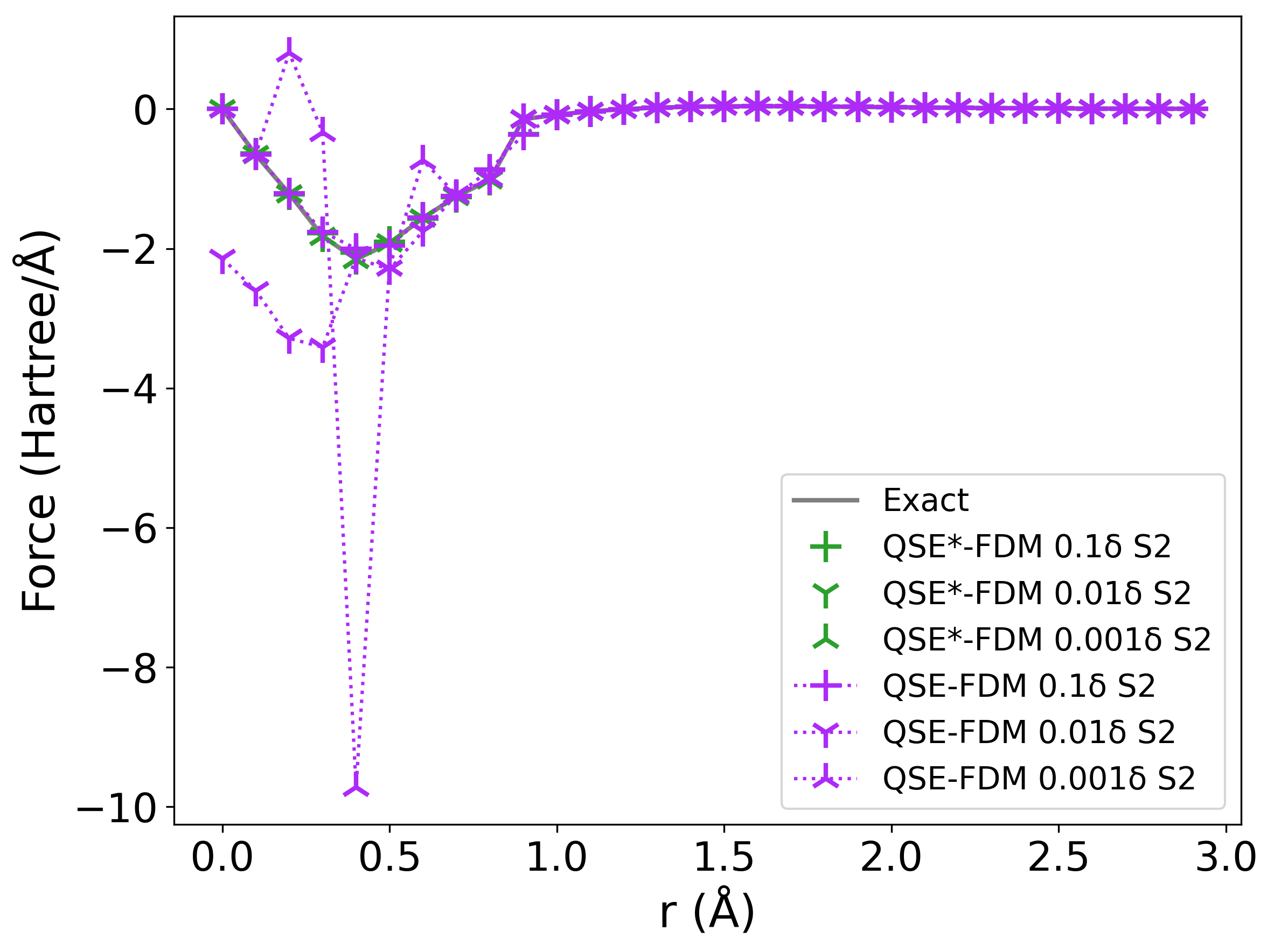}
        \caption{}
    \end{subfigure}
    \caption{HFM force results and FDM force results with different step lengths. (a) QSE*-HFM force of each excited states. (b),(c),(d) FDM force comparison between QSE and QSE* of different FDM step lengths.}
    \label{force_comparisons}
\end{figure}
\begin{table}[h]
    \centering
    \begin{tabular}{lS[table-format=1.2e-2]S[table-format=1.2e-2]S[table-format=1.2e-2]S[table-format=1.2e-2]S[table-format=1.2e-2]}
        \toprule
        $\Delta$F (Ha/Å) & {VQE-HFM S$_0$} & {QSE-HFM S$_1$} & {QSE*-HFM S$_1$} & {QSE-HFM S$_2$} & {QSE*-HFM S$_2$} \\
        \midrule
        RMSE & \num{9.53e-8} & \num{1.96e-2} & \num{2.00e-10} & \num{6.59e-3} & \num{0.00e0} \\
        Max Error & \num{9.60e-7} & \num{7.97e-2} & \num{2.10e-9} & \num{3.80e-2} & \num{0.00e0} \\
        MAE & \num{2.37e-8} & \num{1.05e-2} & \num{0.00e0} & \num{1.94e-3} & \num{0.00e0} \\
        \bottomrule
    \end{tabular}
    \caption{Comparison of $y$-axis force error for HFM methods on different states (S$_0$, S$_1$, S$_2$) for the middle H atom of H$_3^+$.}
    \label{tab:hfm_error_comparison}
\end{table}
\begin{table}[h]
    \centering
\begin{tabular}{lS[table-format=1.2e-1]S[table-format=1.2e-1]S[table-format=1.2e-1]S[table-format=1.2e-1]S[table-format=1.2e-1]}
\toprule
$\Delta$F (Ha/Å) & {VQE-FDM S$_0$} & {QSE-FDM S$_1$} & {QSE*-FDM S$_1$} & {QSE-FDM S$_2$} & {QSE*-FDM S$_2$} \\
\midrule
\multicolumn{6}{c}{$\delta = 0.001$} \\
RMSE & \num{1.23e-6} & \num{4.74e0} & \num{4.28e-6} & \num{2.81e-1} & \num{4.32e-6} \\
Max Error & \num{9.32e-6} & \num{2.56e1} & \num{3.44e-5} & \num{1.78e0} & \num{3.51e-5} \\
MAE & \num{3.62e-7} & \num{1.47e0} & \num{1.13e-6} & \num{8.90e-2} & \num{1.24e-6} \\
\midrule
\multicolumn{6}{c}{$\delta = 0.010$} \\
RMSE & \num{1.23e-4} & \num{6.09e-1} & \num{4.27e-4} & \num{2.37e-1} & \num{4.31e-4} \\
Max Error & \num{9.33e-4} & \num{2.55e0} & \num{3.41e-3} & \num{1.36e0} & \num{3.47e-3} \\
MAE & \num{3.62e-5} & \num{2.64e-1} & \num{1.13e-4} & \num{8.04e-2} & \num{1.24e-4} \\
\midrule
\multicolumn{6}{c}{$\delta = 0.100$} \\
RMSE & \num{1.27e-2} & \num{2.41e-1} & \num{3.15e-2} & \num{2.32e-1} & \num{3.21e-2} \\
Max Error & \num{9.73e-2} & \num{1.31e0} & \num{2.02e-1} & \num{1.37e0} & \num{2.13e-1} \\
MAE & \num{3.66e-3} & \num{9.40e-2} & \num{9.95e-3} & \num{8.15e-2} & \num{1.11e-2} \\
\bottomrule
\end{tabular}
    \caption{Comparison of $y$-axis force error on the middle H atom of H$_3^+$ with different FDM methods and step lengths.}
    \label{tab:h3pfdm}
\end{table}
The results presented in Figure~\ref{force_comparisons} and Table~\ref{tab:hfm_error_comparison}-\ref{tab:h3pfdm} provide evaluations of the accuracy and robustness of quantum\replaced{-computing electronic structure solver}{methods} for computing nuclear forces. Specifically, we compare the HFM \added{(as in eq.(\ref{eqhfm}))} and FDM \added{(as in eq.(\ref{eqfdm}))}\comment{R3 Q10} applied within VQE and QSE frameworks, including an QSE* variant. These approaches are assessed against exact analytical forces of reference. \added{Since the calculation of H$_3^+$ naturally locates on a plane. Here on the model trajectory (freezing the positions of the first and third hydrogen atoms), we define the left and right hydrogen atoms as being on the $x$-axis, and the middle hydrogen atom as starting from the midpoint and moving to one side along the $y$-axis.}\comment{R3 Q12} We focusing on the $y$-component of the force on the central hydrogen atom as a function of separation distance $r$. 

The HFM results, as illustrated in the noiseless simulations of Figure~\ref{force_comparisons}(a), demonstrate that the QSE*-HFM approach yields forces that closely track the exact curve across all examined states (S$_0$, S$_1$, S$_2$), with minimal deviations even in regions of steep potential gradients or of conical intersection. Quantitative error of forces in Table~\ref{tab:hfm_error_comparison} underscore this fidelity: for the QSE*-HFM method, RMSE are on the order of $10^{-10}$~Ha/Å~or lower for S$_1$ and S$_2$, with max error not exceeding $2.1 \times 10^{-9}$~Ha/Å~and MAE effectively zero within numerical precision. In contrast, QSE-HFM exhibits significantly higher errors for excited states, with RMSE values of $1.96 \times 10^{-2}$~Ha/Å~for S$_1$ and $6.59 \times 10^{-3}$~Ha/Å~for S$_2$, reflecting challenges in capturing subspace instabilities. For the ground state (S$_0$), VQE-HFM achieves sub-microhartree accuracy (Max Error $9.60 \times 10^{-7}$~Ha/Å). These findings highlight the applicability of the QSE*, which incorporates additional operators to replenish subspace, thereby enabling HFM to deliver accurate forces for multi-state dynamics without empirical corrections.

Turning to the FDM in Figure~\ref{force_comparisons} (b), (c), (d), the force profiles reveal a strong dependence on the finite-difference step size $\delta$. For small $\delta$ (e.g., 0.001Å), both QSE-FDM and QSE*-FDM approximate the exact forces well. Table~\ref{tab:h3pfdm} quantifies this trend: at $\delta = 0.001$Å, RMSE for VQE-FDM S$_0$ is $1.23 \times 10^{-6}$~Ha/Å, while QSE-FDM S$_1$ balloons to $4.74$~Ha/Å, indicative of amplified errors. Increasing $\delta$ to 0.01Å~and 0.1Å~systematically degrades accuracy across all methods, with RMSE rising by 2 orders of magnitude.\deleted{as the FDM transitions from truncation-error-dominated to discretization-error-dominated regimes.}\comment{R3 Q13} Notably, QSE* consistently outperforms QSE in FDM contexts, suggesting that the extended subspace better stabilizes finite differences.

\begin{figure}[h]
    \centering
    \begin{subfigure}{0.45\textwidth}
        \includegraphics[width=\textwidth]{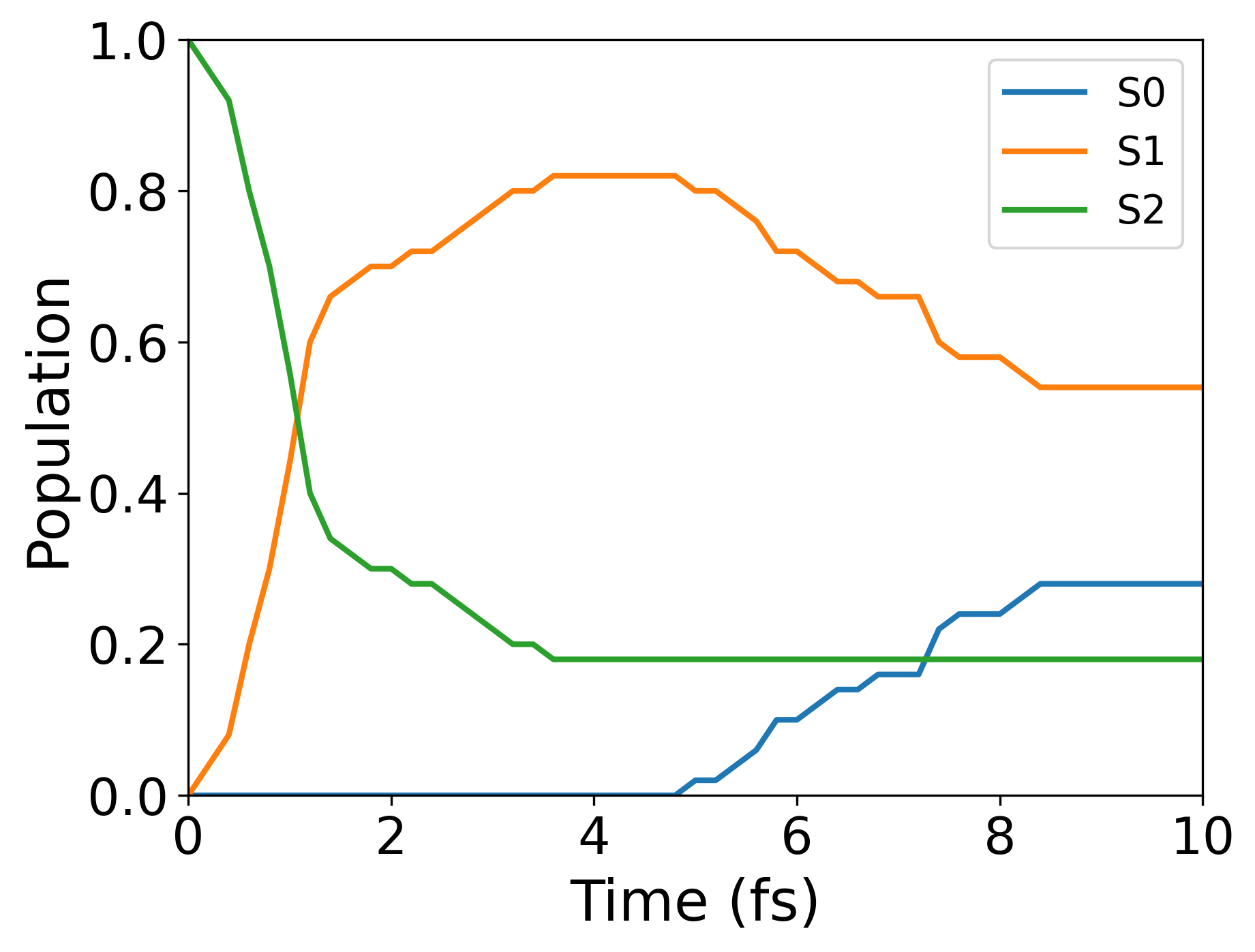}
        \caption{Exact FCI as the electronic structure solver, without curvature-induced hopping correction}
    \end{subfigure}
    \hfill
    \begin{subfigure}{0.45\textwidth}
        \includegraphics[width=\textwidth]{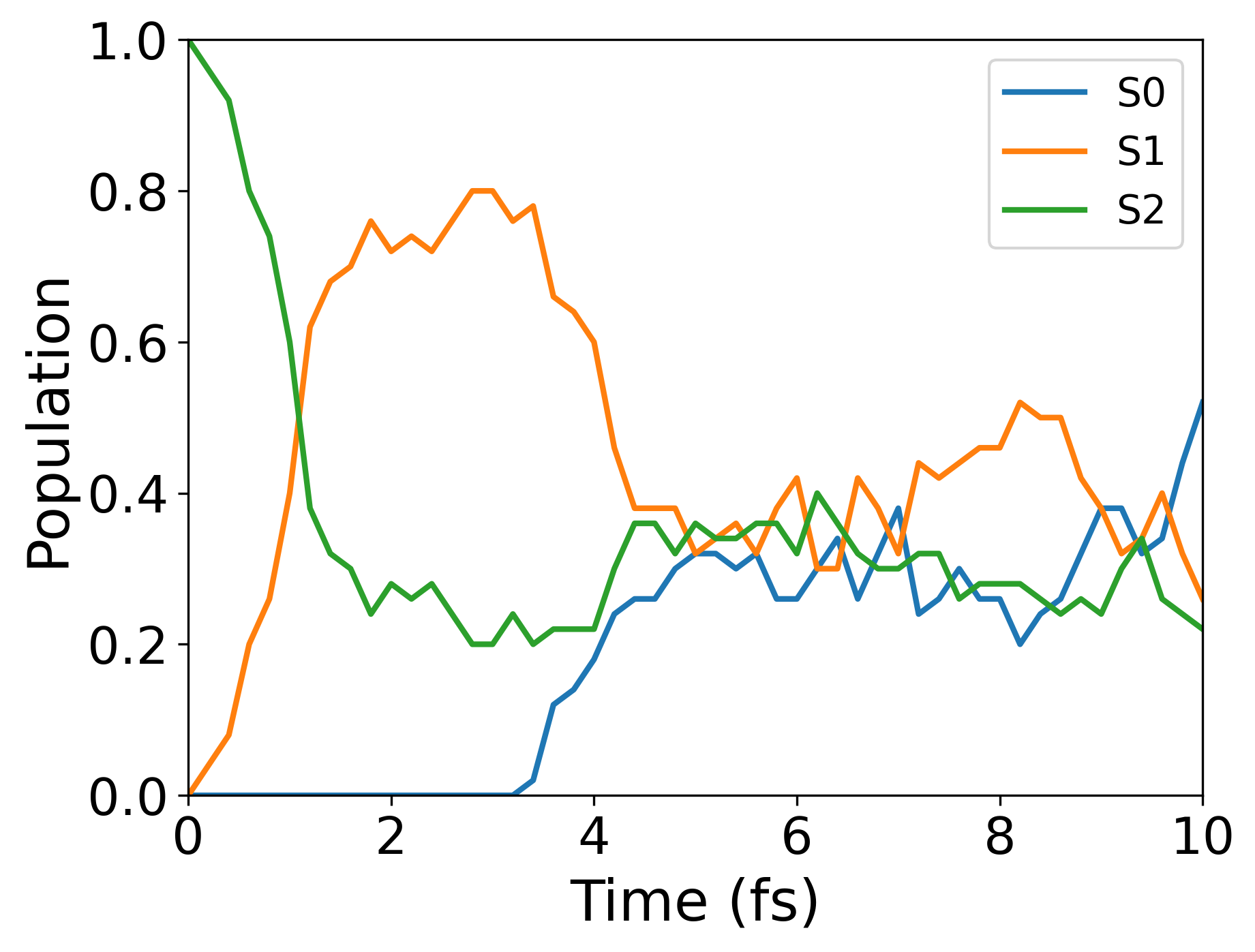}
        \caption{VQE-QSE, without curvature-induced hopping correction}
    \end{subfigure}
      \vspace{0.5cm}
        \begin{subfigure}{0.45\textwidth}
        \includegraphics[width=\textwidth]{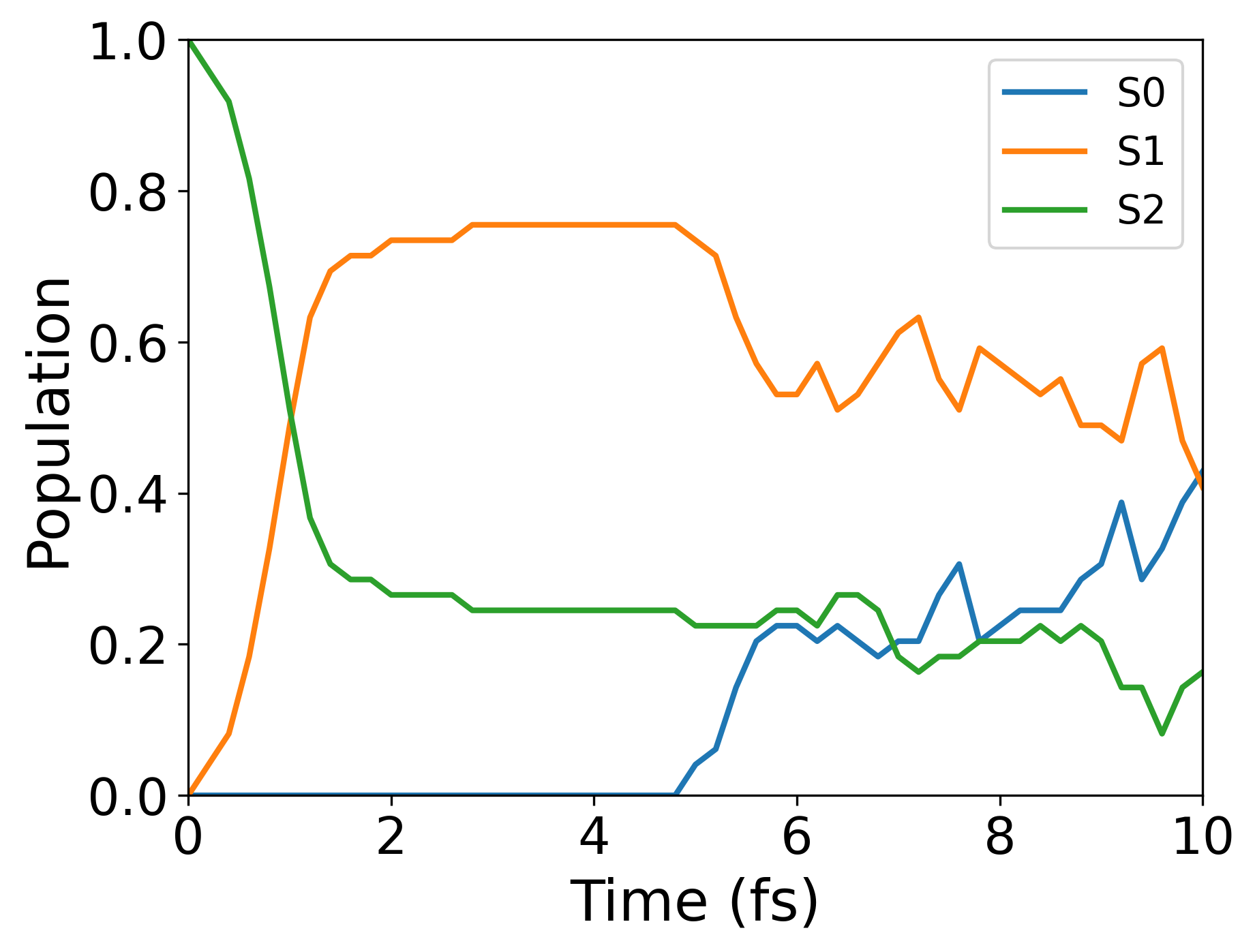}
        \caption{VQE-QSE*, without curvature-induced hopping correction}
    \end{subfigure}
    \hfill
    \begin{subfigure}{0.45\textwidth}
        \includegraphics[width=\textwidth]{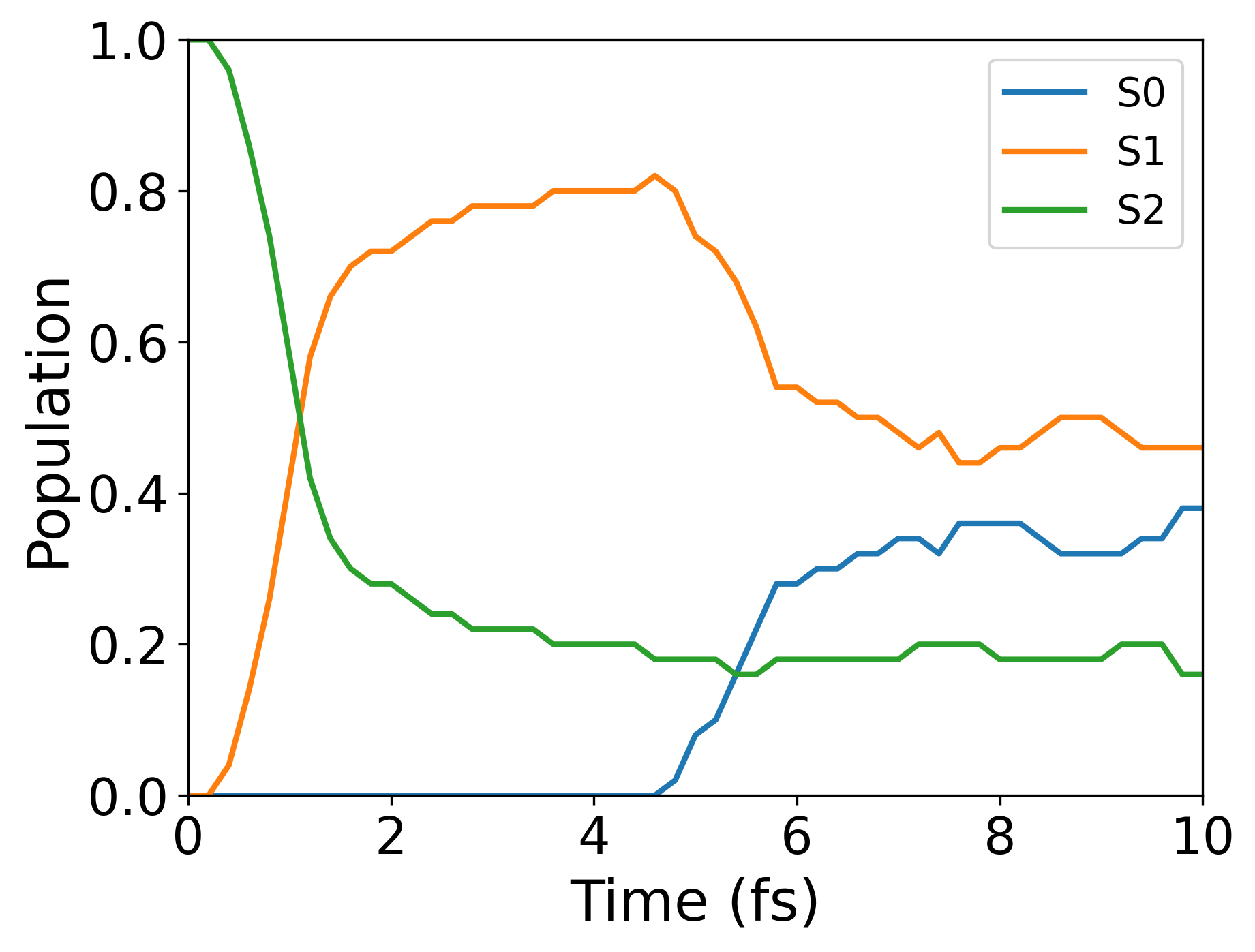}
        \caption{VQE-QSE*, with curvature-induced hopping correction applied}
        \label{popblock}
    \end{subfigure}
    \caption{50-trajectory NAMD population evolution of H$_3^+$ with different electronic structure solver and hopping rules. Initial state at S$_2$.}
     \label{h3ppop1}
\end{figure}
Figure~\ref{h3ppop1} demonstrate the population evolution during the NAMD simulations of the H$_3^+$ dissociation process, comparing the performance of reference classical\added{-computing}\comment{R3 Q3} exact solver with quantum computing approaches, including the VQE, QSE and QSE* method. Figure~\ref{h3ppop1}(a) and Figure~\ref{h3ppop1}(b) illustrates the state populations as a function of time for the reference \deleted{classical} FCI (left\added{, on classical computer}) and VQE-QSE (right\added{, simulated on classical computer via quantum algorithm simulator}\comment{R3 Q3}) methods. In the FCI-driven simulation, which serves as a \deleted{classical} reference, the state S$_1$ (blue) depopulates gradually over the initial 4 fs, transferring population primarily to S$_1$ (orange), which peaks around 5 fs before decaying. After 5 fs, S$_0$ (blue) receives a smaller but steady population increase. This behavior is indicative of efficient non-adiabatic
 transfer driven by conical intersections in the H$_3^+$ PESs, which are well-known to facilitate ultra-fast relaxation in this system\cite{jcph3}. In contrast, the VQE-QSE based simulation exhibits more oscillatory population transfers, with S$_1$ showing pronounced fluctuations between 2 and 6 fs, and S$_2$ displaying erratic rises and falls, indicative of PES fractures and instability.

 The population evolution by the VQE-QSE* with canonical LZSH (Figure~\ref{h3ppop1}(c)) and VQE-QSE* with curvature-induce-corrected LZSH (Figure~\ref{h3ppop1}(d)) are presented. Figure~\ref{h3ppop1}(c) behaves smooth transition but instantly turn to oscillations after $6$ fs, suggesting sudden emergence of hopping events at the late stage of dissociation, which we ascribe to the discontinuity of PESs with small displacement interval at the far dissociation plateau, where PESs tend to be close and parallel. In Figure~\ref{h3ppop1}(d), the application of the curvature-driven hopping correction technique \added{(as introduced in eq.(\ref{curvaturecorrection}))}\comment{R3 Q10, R3 Q15} significantly stabilizes the dynamics, makes populations evolve more smoothly, without losing essential physical picture of the evolution, with S$_2$ smooth depopulating, S$_1$ \& S$_0$ stabilizing, closely mimicking the reference population behavior. 

 \added{In addition, the sharp oscillations observed in Figure~\ref{h3ppop1}(b) and \ref{h3ppop1}(c) arise not only from the numerical instability of the quantum-computing PES solver but also from an insufficient number of trajectories. In surface-hopping NAMD simulations, a sufficiently large number of trajectories is essential for mitigating statistical noise and achieving reliable ensemble averaging. However, emulating quantum algorithms on classical computers is computationally demanding; thus, to accommodate limited resources, all simulations presented in Figure~\ref{h3ppop1} employ only 50 trajectories. The erratic fluctuations in panel (b) persist throughout the dynamics, highlighting the inherent instability of the VQE-QSE solver and the inadequacy of the trajectory count. Panel (c), which employs the improved VQE-QSE* method without curvature-induced hopping correction, exhibits smooth initial behavior but deviates from the exact reference. A similar trend is observed in panel (d) upon application of the curvature correction, where the population evolution does not precisely match the reference. These artifacts primarily originate from the insufficient number of trajectories in the LZSH ensemble.}\comment{R3 Q14}

\begin{figure}[h]
    \centering
    \begin{subfigure}{0.45\textwidth}
        \includegraphics[width=\textwidth]{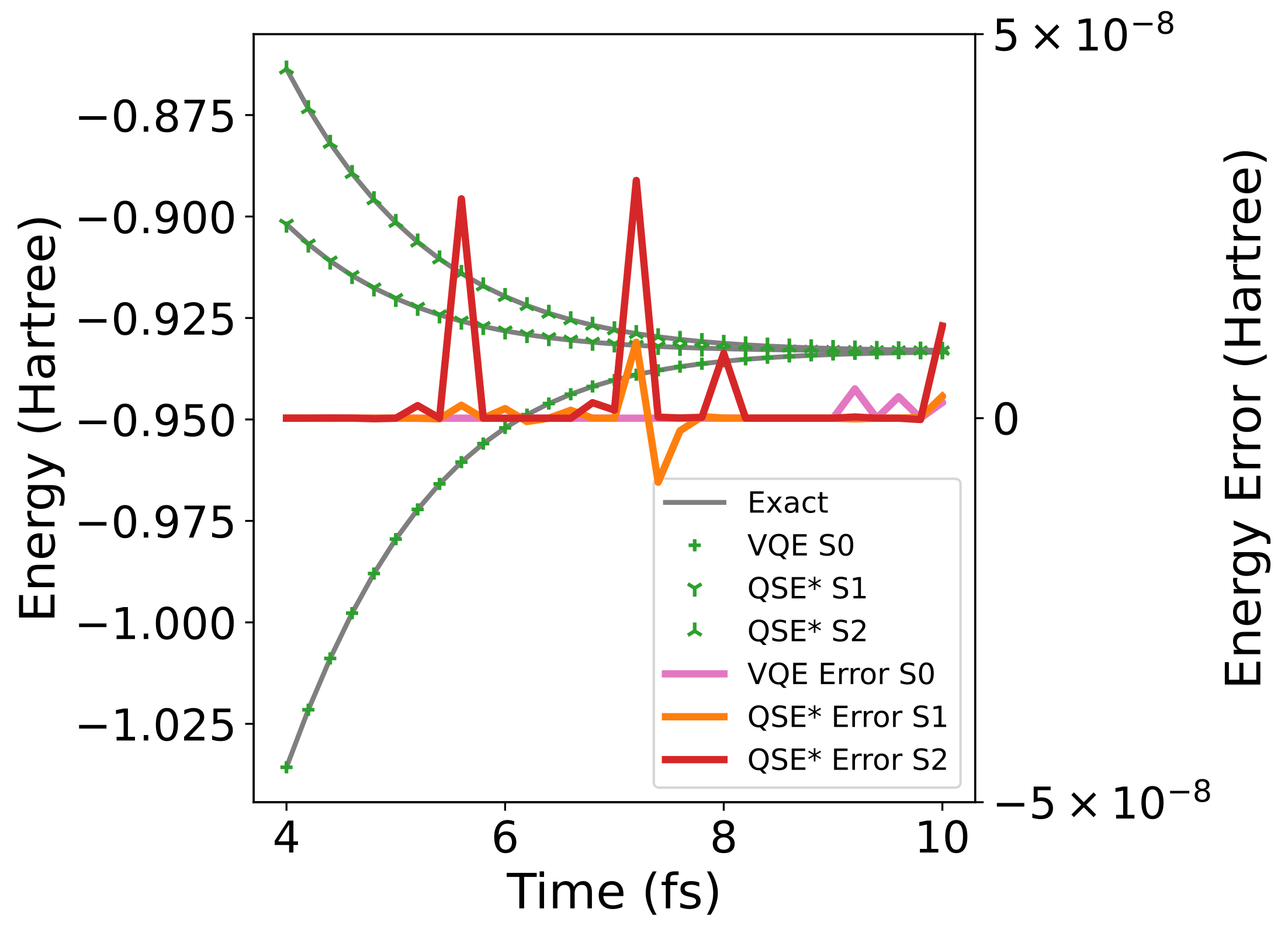}
        \caption{}
    \end{subfigure}
    \hfill
    \begin{subfigure}{0.45\textwidth}
        \includegraphics[width=\textwidth]{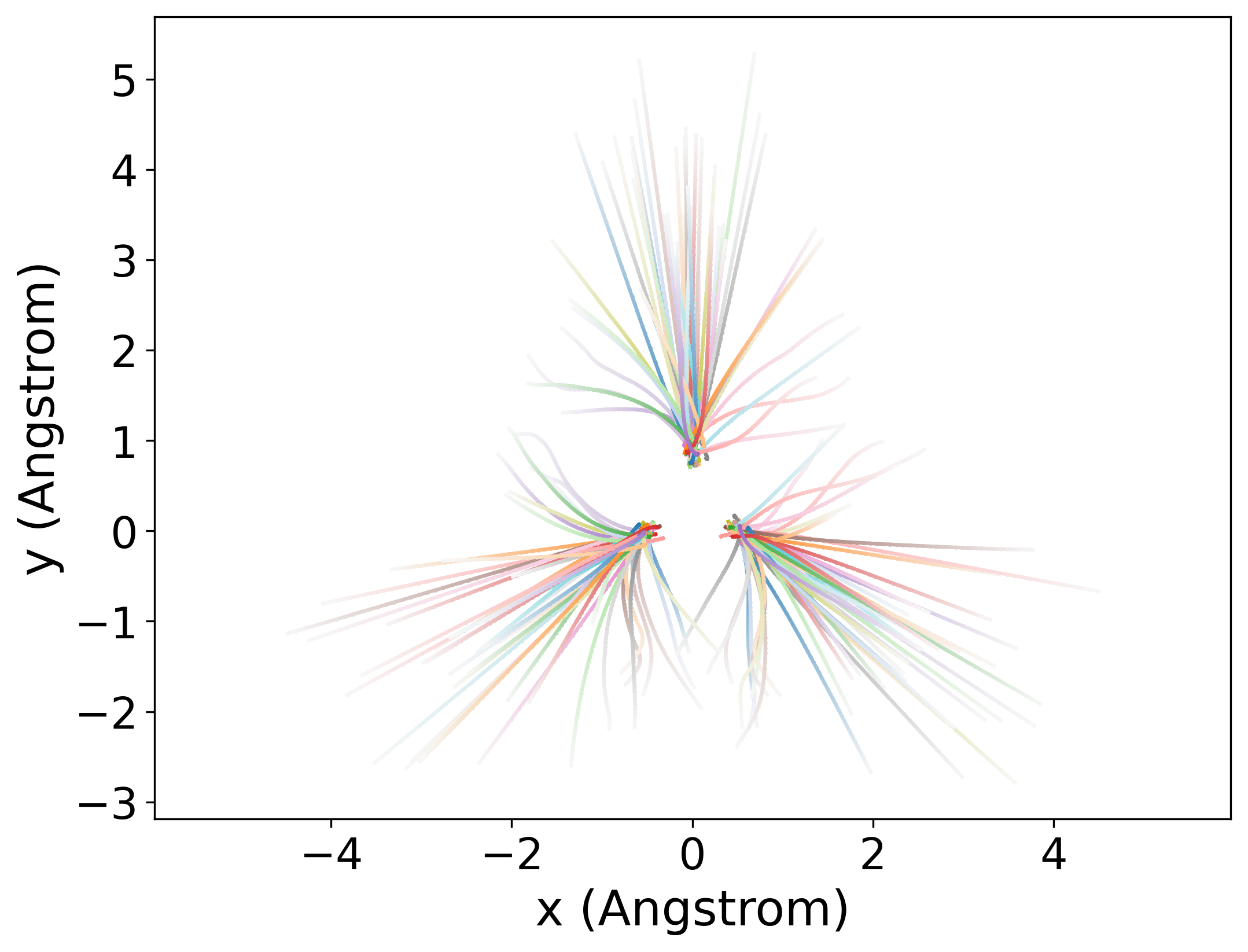}
         \caption{}
    \end{subfigure}
    \caption{(a) PESs at the end of the dissociation and their instability, taking the view that displacement interval of geometry is small. (b) Geometry trajectories of the ensemble during the NAMD simulation.}
    \label{h3pmicrodissaso}
\end{figure}
To quantify the accuracy of the quantum electronic structure solver as well as to reveal the underlying cause of the late-6fs oscillation, Figure~\ref{h3pmicrodissaso} (a) compares the PESs at late dissociation times from 4 fs onward, where the surfaces S$_0$, S$_1$, and S$_2$ of both reference and VQE-QSE* converge smoothly and closely. Dot lines depicts the absolute energy errors relative to the exact solution, revealing that VQE-QSE* errors remain below $10^{-7}$ Hartree but spike intermittently. Note that initial guess heritage of VQE during the PES calculation could not fully smoothen such microscopic spike. These artifacts, though small, can induce unphysical hops in regions of near-degeneracy.

The geometric trajectories of the molecular ensemble, visualized in Figure~\ref{h3pmicrodissaso} (b), demonstrate the realistic dissociation trajectories in the x-y plane, given by the hopping corrected LZSH-NAMD with VQE-QSE* solver. The trajectories fan out symmetrically from the central equilibrium geometry, with clusters branching toward positive and negative y-directions. The multi-colored lines indicate temporal evolution, with earlier times near the origin and later dispersion. Atoms in same trajectory shares the same color, whose initial positions and momentum are determined by Wigner sampling.

\subsection{Use case II: C$_2$H$_4$ results in CAS(2,2)}

\begin{figure}[h!]
  \centering
  \begin{subfigure}[b]{0.46\textwidth}
    \centering
    \includegraphics[width=\textwidth]{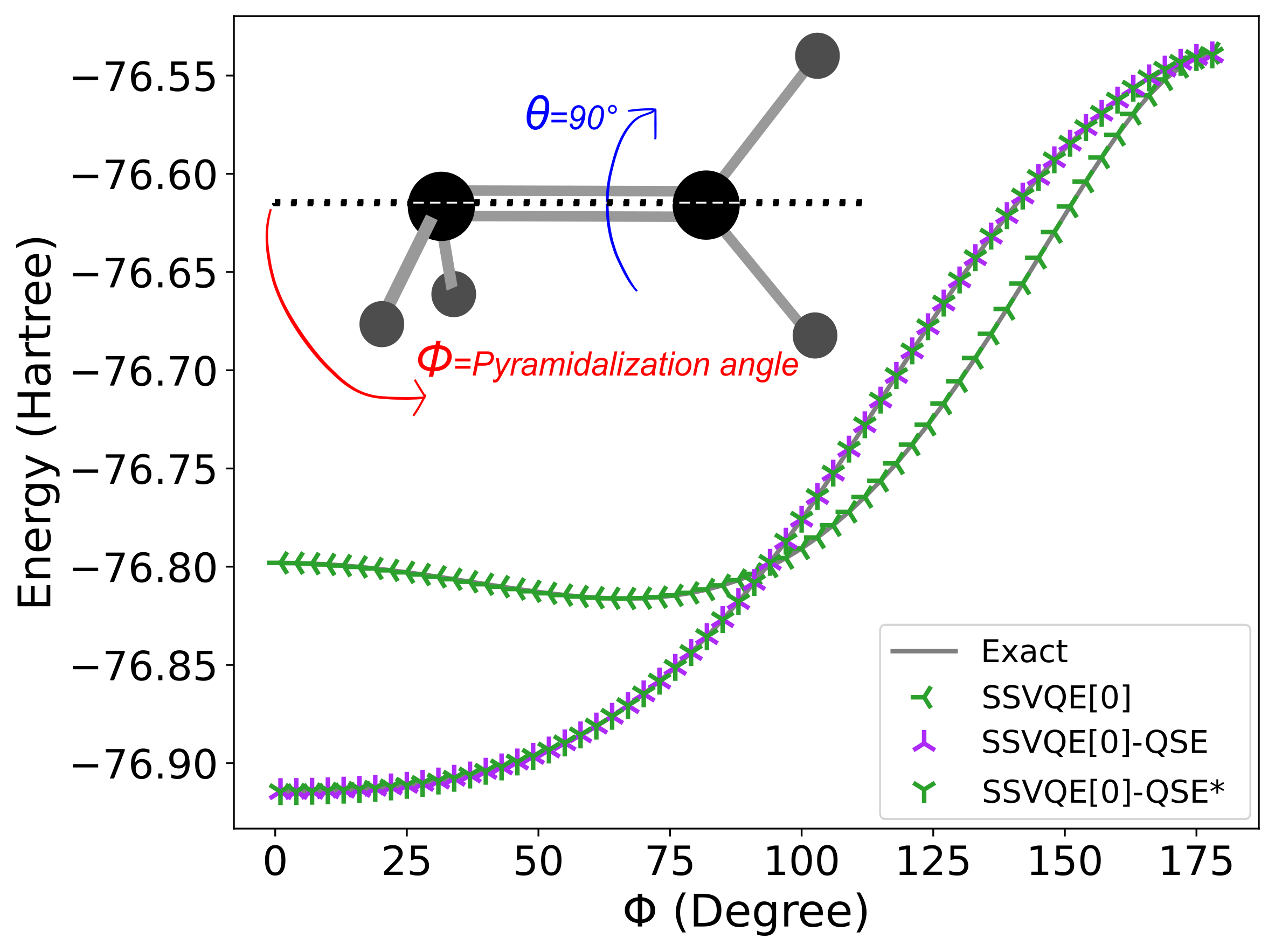}
\caption{}
  \end{subfigure}
  \hfill
  \begin{subfigure}[b]{0.46\textwidth}
    \centering
    \includegraphics[width=\textwidth]{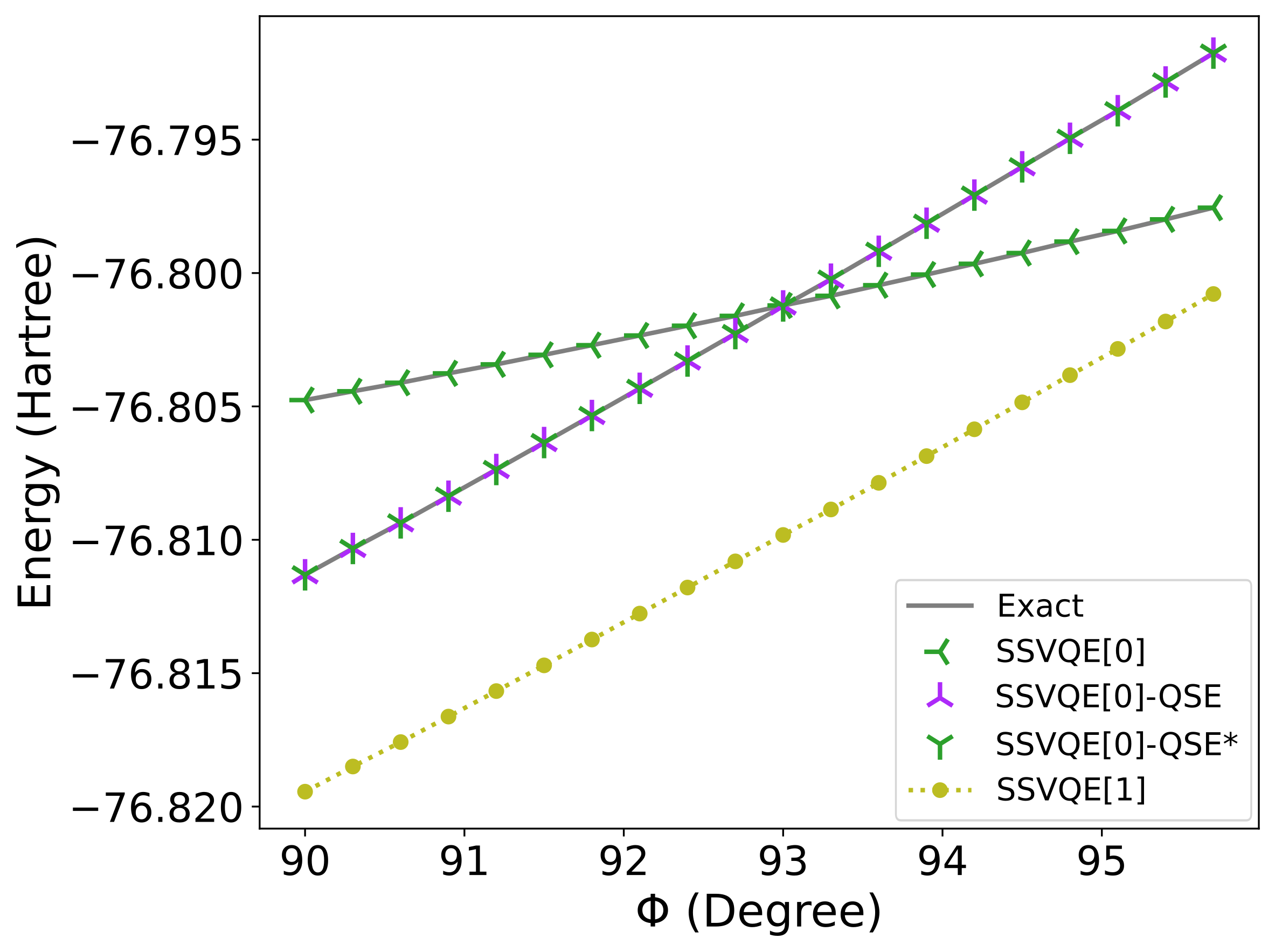}
\caption{}
  \end{subfigure}

  \caption{PES results along demonstrative pyramidalization geometries of C$_2$H$_4$ in CAS(2,2). Gray lines represent the reference results, while markers denote those obtained via quantum electronic structure solvers. (a) PES result with 3 degree's rotation angle interval between data points, SSVQE$\lbrack$0$\rbrack$-QSE indicates the singlet-adapted QSE basing on the reference state given by SSVQE with UCCSD Ansatz in the active space, with only single-double excitation operators used to expand subspaces. SSVQE$\lbrack$0$\rbrack$-QSE* indicates the singlet-adapted QSE* with single and double de-excitation expanding, expanding subspaces upon the reference state searched by SSVQE. (b) PESs around the conical intersection region with data points' rotation angle interval of 0.3 degree, also demonstrate comparison between SSVQE-QSE(QSE*) integrated methods and SSVQE-solo method.}
\label{fig:c2h4_pes}
\end{figure}

\begin{table}[h]
\centering
\caption{Comparison of Macroscopic and Microscopic C$_2$H$_4$ PESs errors for different solvers.}
\begin{tabular}{lcccc}
\toprule
$\Delta$E (Hartree) & SSVQE[0] & SSVQE[0]-QSE & SSVQE[0]-QSE* & SSVQE[1] \\
\midrule
\multicolumn{5}{c}{3-degree interval angle} \\
RMSE & \num{1.033e-13} & \num{7.666e-14} & \num{7.415e-13} & - \\
Max Error & \num{2.416e-13} & \num{1.847e-13} & \num{9.948e-13} & - \\
MAE & \num{8.266e-14} & \num{5.921e-14} & \num{7.375e-13} & - \\
\midrule
\multicolumn{5}{c}{0.3-degree interval angle} \\
RMSE & \num{1.033e-13} & \num{7.666e-14} & \num{7.415e-13} & \num{8.67e-03} \\
Max Error & \num{2.416e-13} & \num{1.847e-13} & \num{9.948e-13} & \num{9.03e-03} \\
MAE & \num{8.266e-14} & \num{5.921e-14} & \num{7.375e-13} & \num{8.60e-03} \\
\bottomrule
\end{tabular}
\label{c2h4peserror}
\end{table}

To assess the adaptablity of the subspace-based quantum electronic structure solvers in capturing the PESs, we examine the pyramidalization pathway of C$_2$H$_4$ within the complete active space (CAS(2,2)) framework, which encompasses the $\pi$ and $\pi^*$ orbitals. Figure~\ref{fig:c2h4_pes} (a) illustrates the PES with large displacement interval along the pyramidalization angle $\phi$, comparing the exact diagonalization results (gray lines) with SSVQE[0]-QSE and SSVQE[0]-QSE*. 
 
Quantitative energy errors for these solvers are summarized in Table~\ref{c2h4peserror}. Among PESs with an interval degree of $3$ and $0.3$ between data points, all the hybrid subspace-based solvers reached sub-microhartree accuracy. Compared to the CAS(3,2) ionic system examined in the preceding section, the current system features a smaller active space, enabling QSE to exhibit high accuracy in the exemplified regime as well. This underscores the utility of QSE in certain scenarios. However, the QSE* demonstrate slightly lower fidelity because expanding subspaces with extra operators would introduce more linear independency, which lead to larger condition number of overlapping matrix $S$, affect numerical stability.

Here, we present the results prior to energy ordering of the states. Unlike the conventional senario, where VQE is used to search the ground state followed by QSE expansion to obtain excited states, here, SSVQE \added{(as introduced in eq.(\ref{eq:ssvqe_cost})} yields the higher-energy $V$ state\cite{yixi} in the region before the conical intersection. In contrast, QSE, by expanding the subspace, expanding the lower-energy $N$ state in this region. Furthermore, SSVQE demonstrates robust state-tracking capabilities. After the conical intersection, SSVQE continues to track the $V$ state (which now becomes the ground state), while QSE, leveraging the reference state from this region, accurately extends to the higher-energy excited state.

A more detailed examination of the region of PESs with an interval of $0.3$ degree around the conical intersection is provided in Figure~\ref{fig:c2h4_pes}(b), where we compare the energy of N-state and V-state directly from SSVQE. The hybrid subspace-based solvers maintain high fidelity to the exact curves, with correct transitions through the conical intersection region. In contrast, the SSVQE-solo results deviate noticeably. This behavior indicates that where the weighted sum of energies from orthogonal references (Eq.~\ref{eq:ssvqe_cost}) is insufficient to resolve all subspaces without additional constraints, requies more advanced SSVQE extension or other methods.

\begin{figure}
    \centering
  \centering
  \begin{subfigure}[b]{0.46\textwidth}
    \centering
    \includegraphics[width=\textwidth]{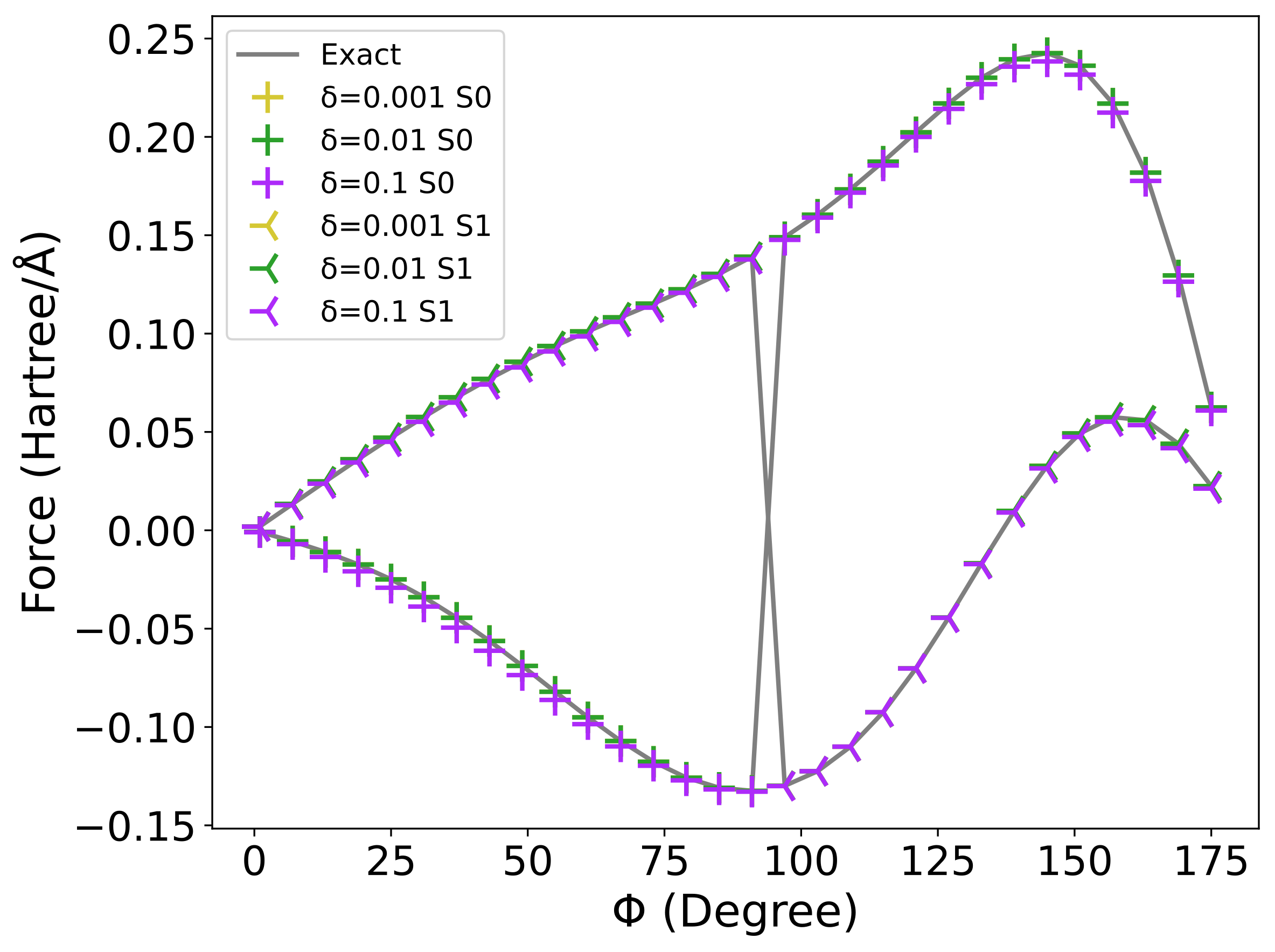}
\caption{}
  \end{subfigure}
  \hfill
  \begin{subfigure}[b]{0.46\textwidth}
    \centering
    \includegraphics[width=\textwidth]{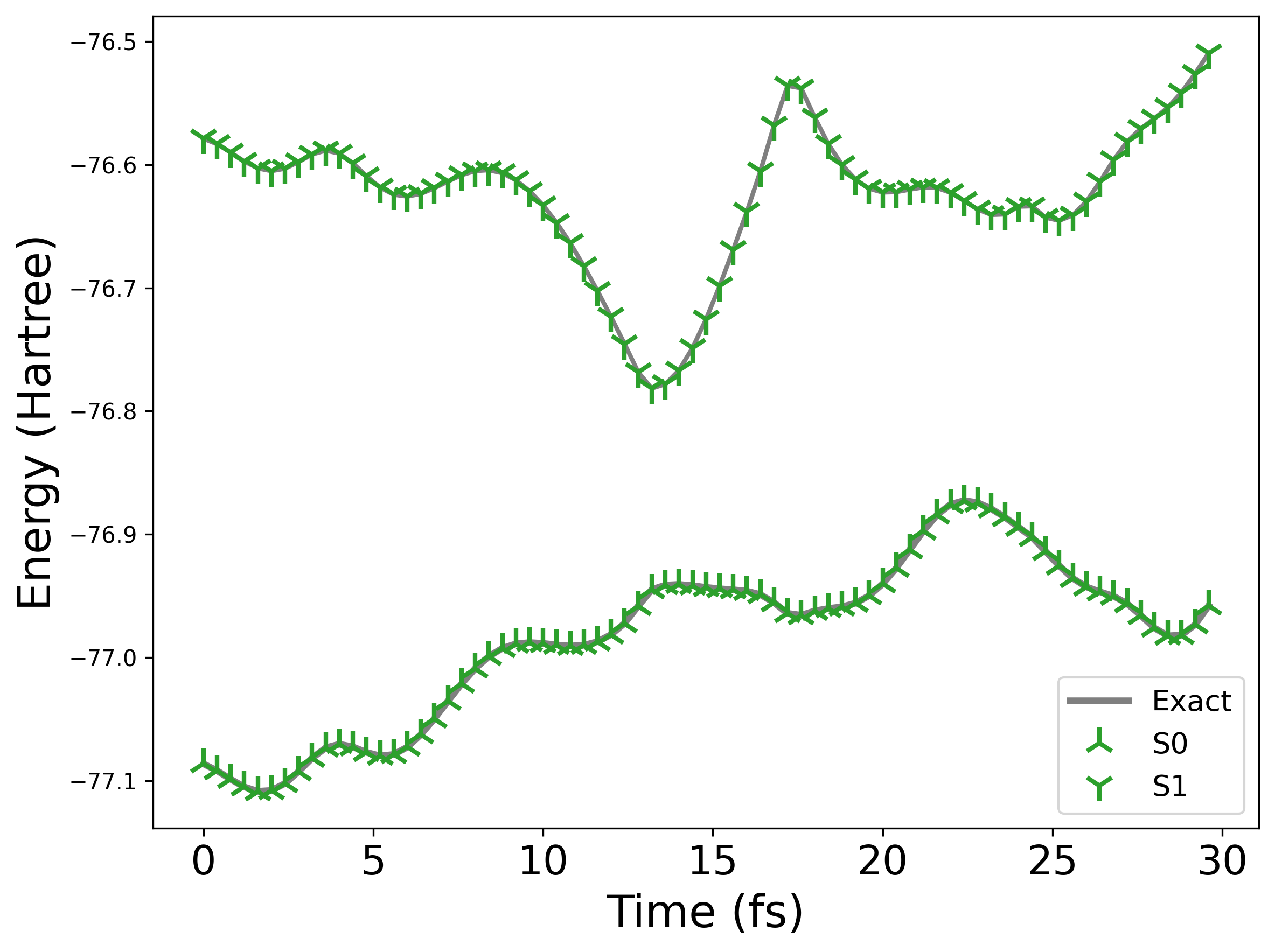}
\caption{}
\end{subfigure}

    \caption{Electronic-structure properties results necessary for NAMD. (a) $x-$axis Force results along demonstrative pyramidalization geometries of the first carbon atom of C$_2$H$_4$ in CAS(2,2). Gray lines represent the reference results, while markers of different colors denote those obtained via SSVQE-QSE with different FDM steplengths. (b) PES result by SSVQE-QSE* along an reference LZSH-NAMD trajectory, 30fs, 75 steps.}
    \label{fig:c2h4noiselessfdm}
\end{figure}
\begin{table}[h]
\centering
\caption{Comparison of force error on the first carbon atom of C$_2$H$_4$ with different FDM steps.}
\label{tab:error_metricsc2h4}
\begin{tabular}{lS[table-format=1.10e-1]S[table-format=1.10e-1]}
\toprule
$\Delta$F (Hartree/Angstrom) & {S$_0$} & {S$_1$} \\
\midrule
\multicolumn{3}{c}{$\delta = \num{0.001}$ Hatree} \\
RMSE & \num{3.740e-7} & \num{2.187e-7} \\
Max Error & \num{6.312e-7} & \num{4.209e-7} \\
Mean Absolute Error & \num{3.238e-7} & \num{1.946e-7} \\
\midrule
\multicolumn{3}{c}{$\delta = \num{0.010}$ Hatree} \\
RMSE & \num{3.29165e-5} & \num{1.78885e-5} \\
Max Error & \num{5.12144e-5} & \num{2.91406e-5} \\
Mean Absolute Error & \num{2.94293e-5} & \num{1.48558e-5} \\
\midrule
\multicolumn{3}{c}{$\delta = \num{0.100}$ Hatree} \\
RMSE & \num{3.2648329e-3} & \num{1.7893454e-3} \\
Max Error & \num{5.0488511e-3} & \num{2.9237422e-3} \\
Mean Absolute Error & \num{2.9277802e-3} & \num{1.4836350e-3} \\
\bottomrule
\end{tabular}
\end{table}
\begin{table}[h]
\centering
\caption{Comparison of PES error for C$_2$H$_4$ along the NAMD trajectory.}
\label{tab:error_metrics_qse}
\begin{tabular}{lS[table-format=1.12e-1]S[table-format=1.12e-1]}
\toprule
$\Delta$E (Hartree) & {S$_0$} & {S$_1$} \\
\midrule
RMSE & \num{1.57e-13} & \num{2.5747180e-8} \\
Max Error & \num{7.53e-13} & \num{1.06571761e-7} \\
Mean Absolute Error & \num{1.12e-13} & \num{1.1482544e-8} \\
\bottomrule
\end{tabular}
\end{table}
The force calculations along the pyramidalization coordinate of the first carbon atom in C$_2$H$_4$ within the CAS(2,2) active space reveal the efficacy of FDM integrated with quantum electronic structure solvers for excited-state properties. As illustrated in Figure~\ref{fig:c2h4noiselessfdm} and Table~\ref{tab:error_metricsc2h4}  taking $x$-axis for demonstration, the reference exact forces (gray line) are closely reproduced by both the hybrid subspace-based solver, with deviations becoming more pronounced at larger FDM step lengths ($\delta$). For $\delta = 0.001$, the computed forces overlay nearly identically with the exact profile across the full range of dihedral angles, capturing changes around the conical intersection. In contrast, larger steps ($\delta = 0.01$ and 0.1) introduce systematic errors, manifesting as offsets.

Complementing the static analysis, Figure~\ref{fig:c2h4noiselessfdm} (b) and Table~\ref{tab:error_metrics_qse} presents the SSVQE-QSE* PESs along the geometry of a LZSH-NAMD reference trajectory, capturing the temporal evolution of S$_0$ and S$_1$ energies in 30 fs. However, along this realistic trajectory, QSE fails largely at many geometries, thus only QSE* results are presented. The SSVQE-QSE* energies for S$_0$ and S$_1$ faithfully reproduce the result by exact reference electronic solver.

\subsection{Noisy Results}
\begin{figure}[h!]
  \centering
  \begin{subfigure}[b]{0.46\textwidth}
    \centering
    \includegraphics[width=\textwidth]{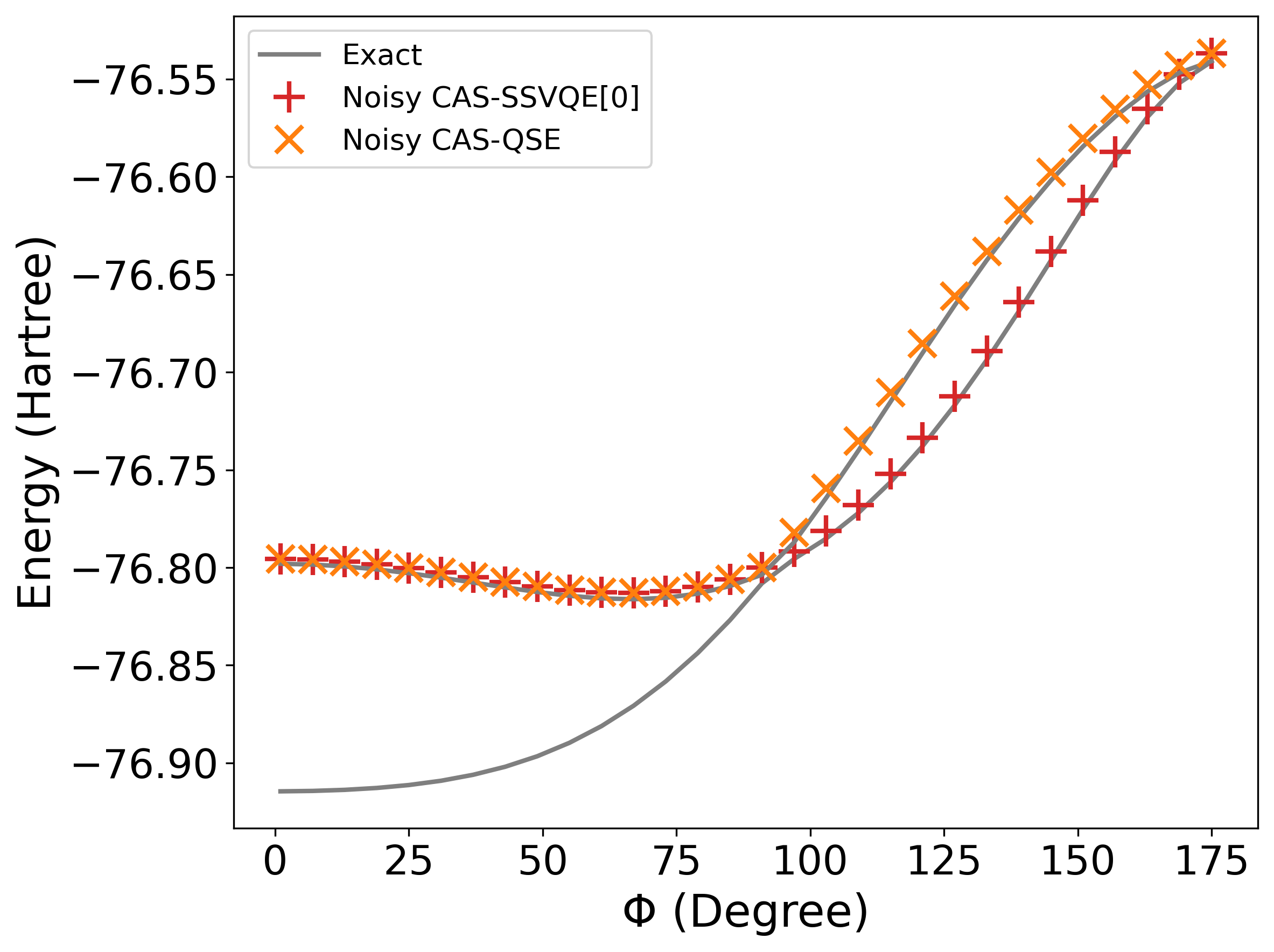}
\caption{}
  
  \end{subfigure}
  \hfill
  \begin{subfigure}[b]{0.46\textwidth}
    \centering
    \includegraphics[width=\textwidth]{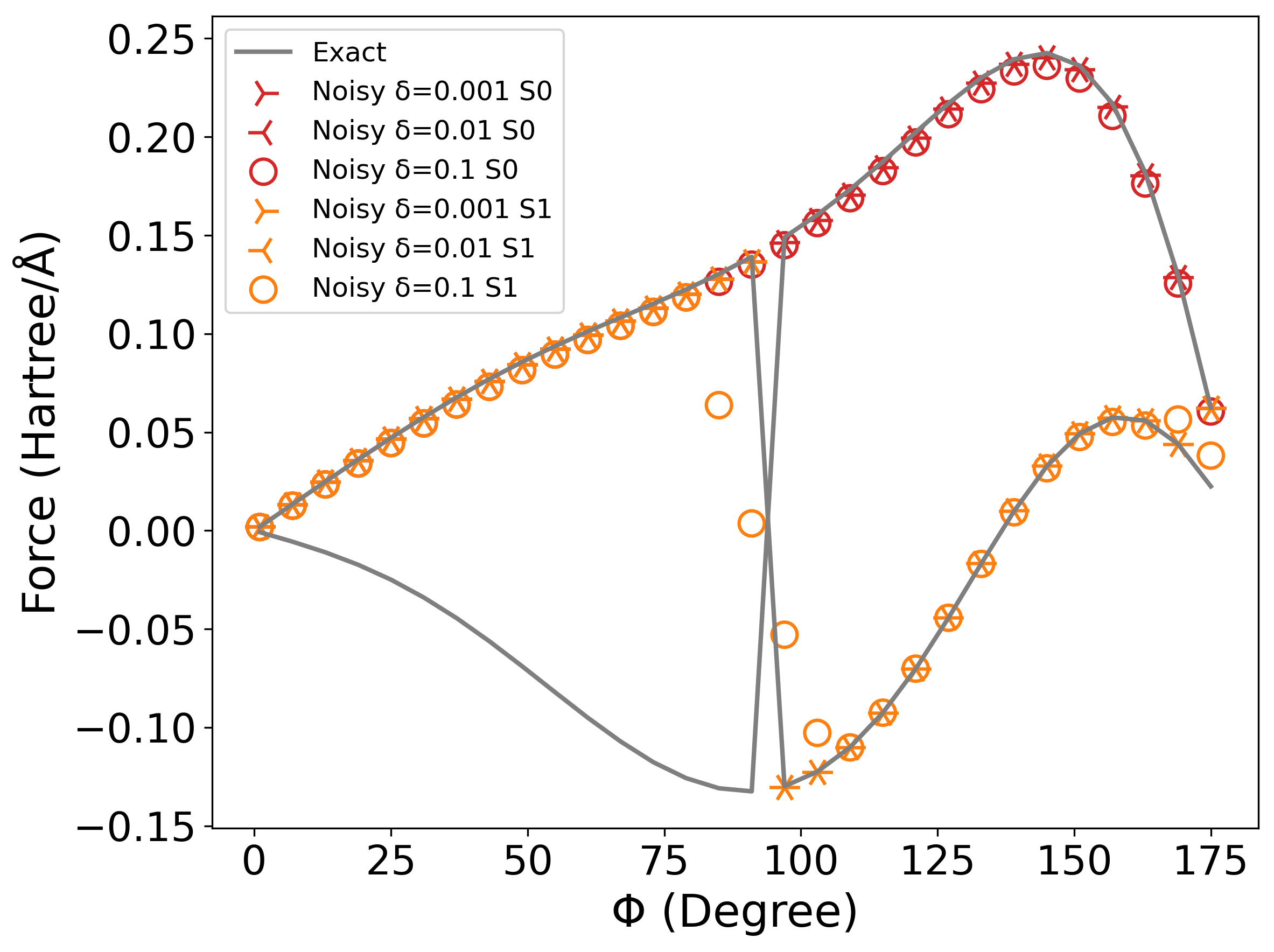}
\caption{}
  \end{subfigure}
\vspace{0.5cm}
  \begin{subfigure}[b]{0.46\textwidth}
    \centering
    \includegraphics[width=\textwidth]{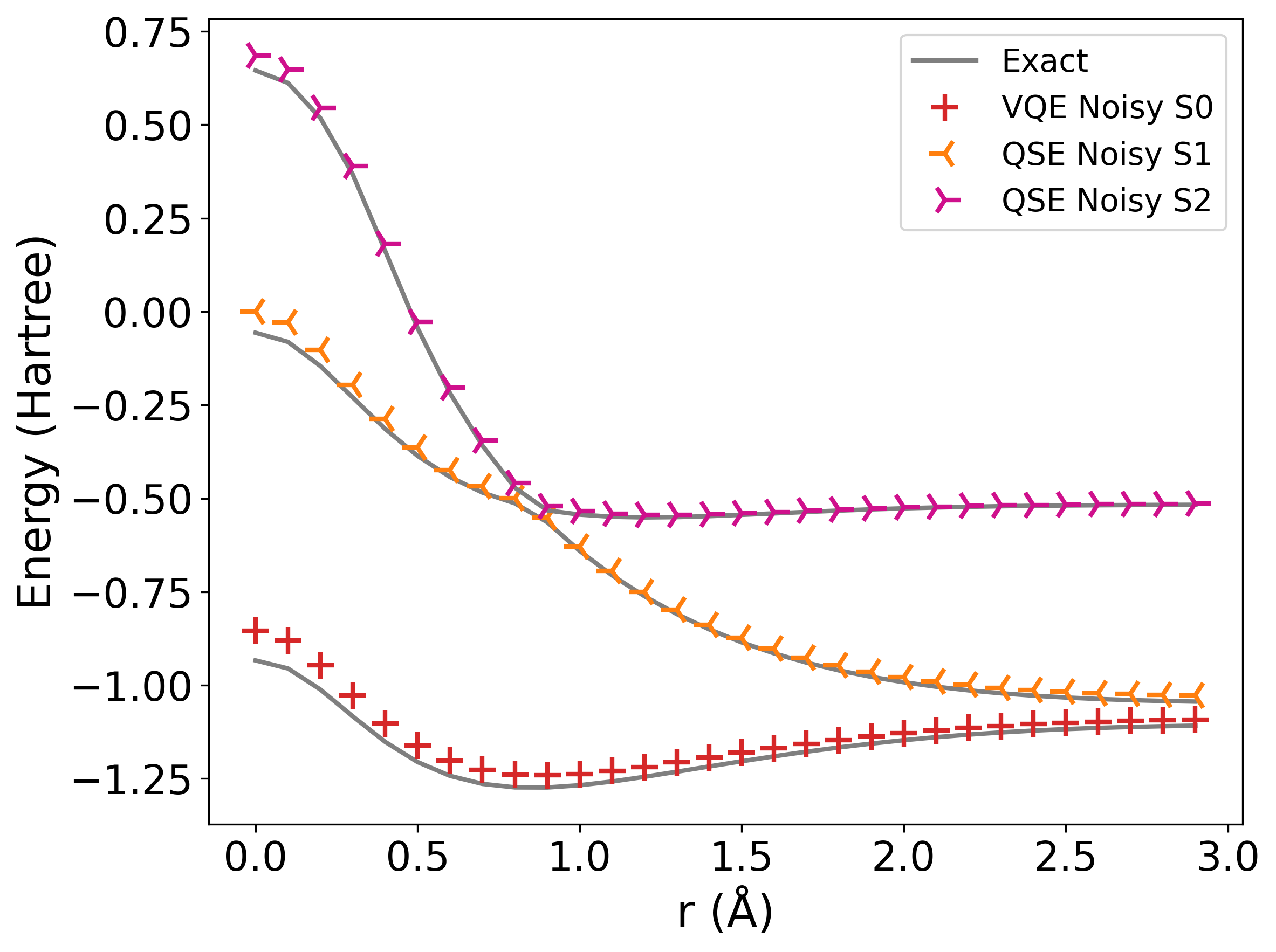}
\caption{}

  \end{subfigure}
  \hfill
  \begin{subfigure}[b]{0.46\textwidth}
    \centering
    \includegraphics[width=\textwidth]{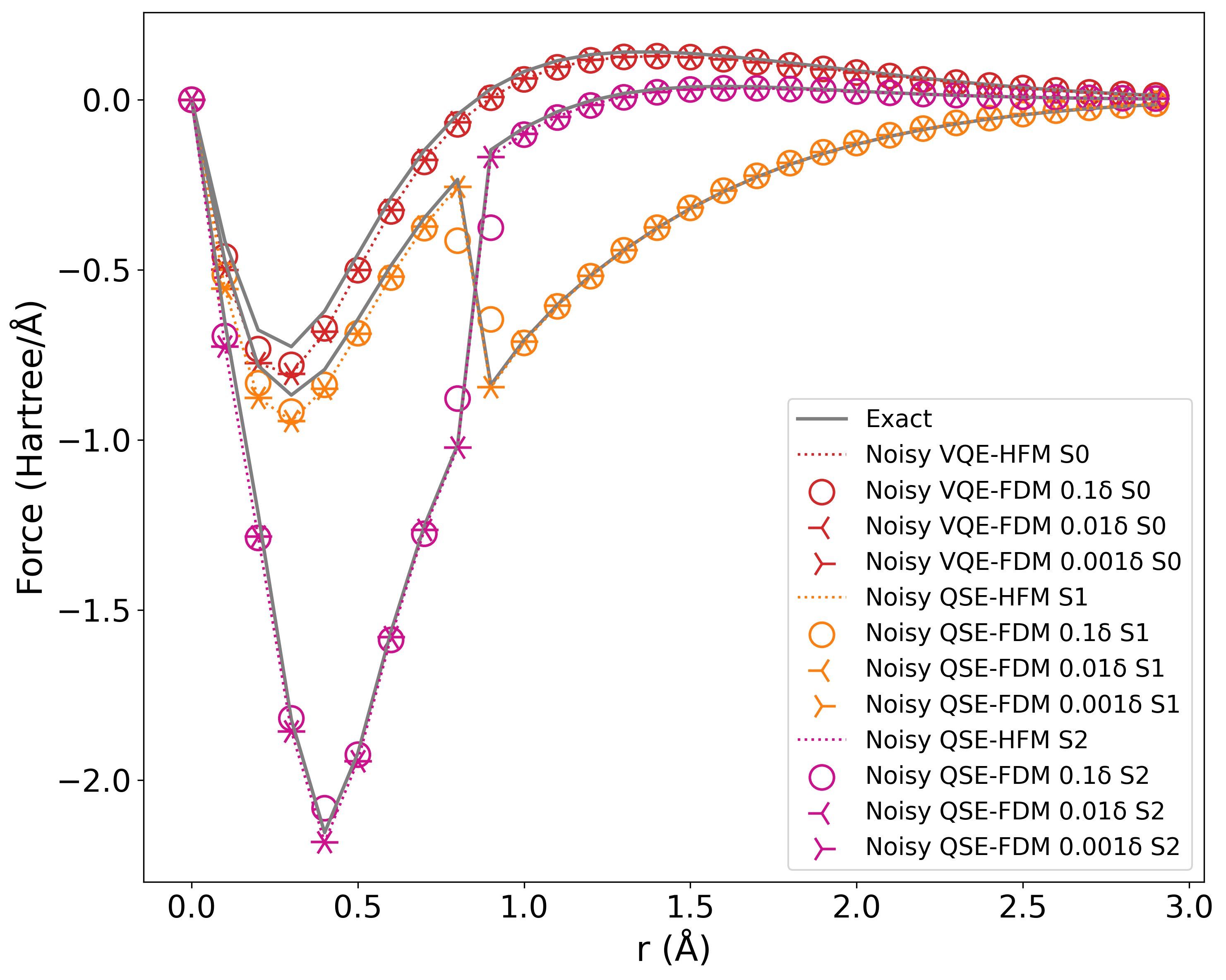}
\caption{}
    
  \end{subfigure}

  \caption{Noisy PES and force result of C$_2$H$_4$ in CAS(2,2) and H$_3^+$ in CAS(3,2). Gray lines represent the reference results, while markers denote those obtained via noisy quantum algorithm simulator on classical computer. (a) Noisy PES result of C$_2$H$_4$. (b)Noisy FDM force result with different step length of C$_2$H$_4$. (c) Noisy PES result of H$_3^+$. (d)Noisy FDM and HFM force results with different step lengths of H$_3^+$. We add depolarization noise with probability of 0.01 for double-qubit gate and 0.001 for single-qubit gate.}
\label{fig:c2h4_pes_noisy}
\end{figure}
\begin{table}[h]
\centering
\caption{Comparison of noisy PES errors of H$_3^+$.}
\label{tab:pes_error_metrics_noisy}
\begin{tabular}{lS[table-format=1.3e-2,table-number-alignment=center]S[table-format=1.3e-2,table-number-alignment=center]S[table-format=1.3e-2,table-number-alignment=center]}
\toprule
$\Delta$E (Hartree) & {S$_0$} & {S$_1$} & {S$_2$} \\
\midrule
RMSE      & \num{3.556e-2} & \num{2.240e-2} & \num{1.409e-2} \\
Max Error & \num{7.966e-2} & \num{5.650e-2} & \num{3.961e-2} \\
MAE       & \num{3.092e-2} & \num{1.917e-2} & \num{9.822e-3} \\
\bottomrule
\end{tabular}
\end{table}

In Figure~\ref{fig:c2h4_pes_noisy} and \added{Table~\ref{tab:pes_error_metrics_noisy}} we demonstrate the PES and nuclear forces of two chemical systems C$_2$H$_4$ and H$_3^+$, under noisy quantum simulation conditions. For C$_2$H$_4$, in the range before the conical intersection, the noisy QSE fails to expand the $N$-state basing on the $V$-state searched by noisy SSVQE. This revealed that quantum noise significantly impacts the cooperation of the hybrid subspace-based quantum solver, preventing it from reaching the target states. 

In contrast, the H$_3^+$ system exhibits a smoother response to noise, manifesting as a consistent offset in the PES and nuclear forces. Notably, QSE calculations based on noisy ground states yielded excited-state deviations smaller than those of the corresponding noisy ground states. This behavior is attributed to the error mitigation capabilities inherent in the QSE framework\added{ (as introduced in the previous section)}\comment{R3 Q18}, show its robustness in noisy environments.

\subsection{Triplet Results}
\begin{figure}[h!]
  \centering
  \begin{subfigure}[b]{0.46\textwidth}
    \centering
    \includegraphics[width=\textwidth]{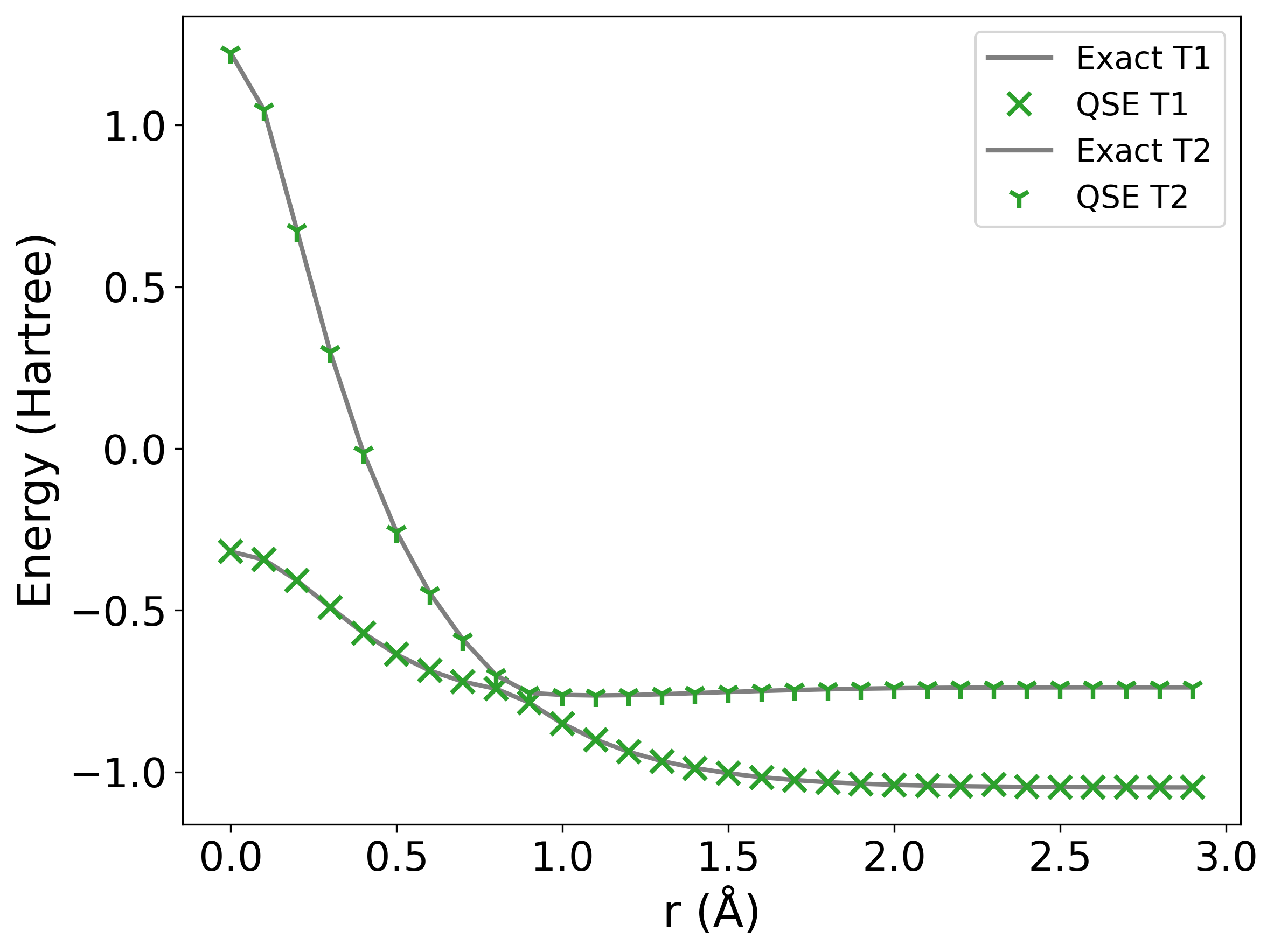}
\caption{}
  \end{subfigure}    
  \hfill
    \begin{subfigure}[b]{0.46\textwidth}
    \centering
    \includegraphics[width=\textwidth]{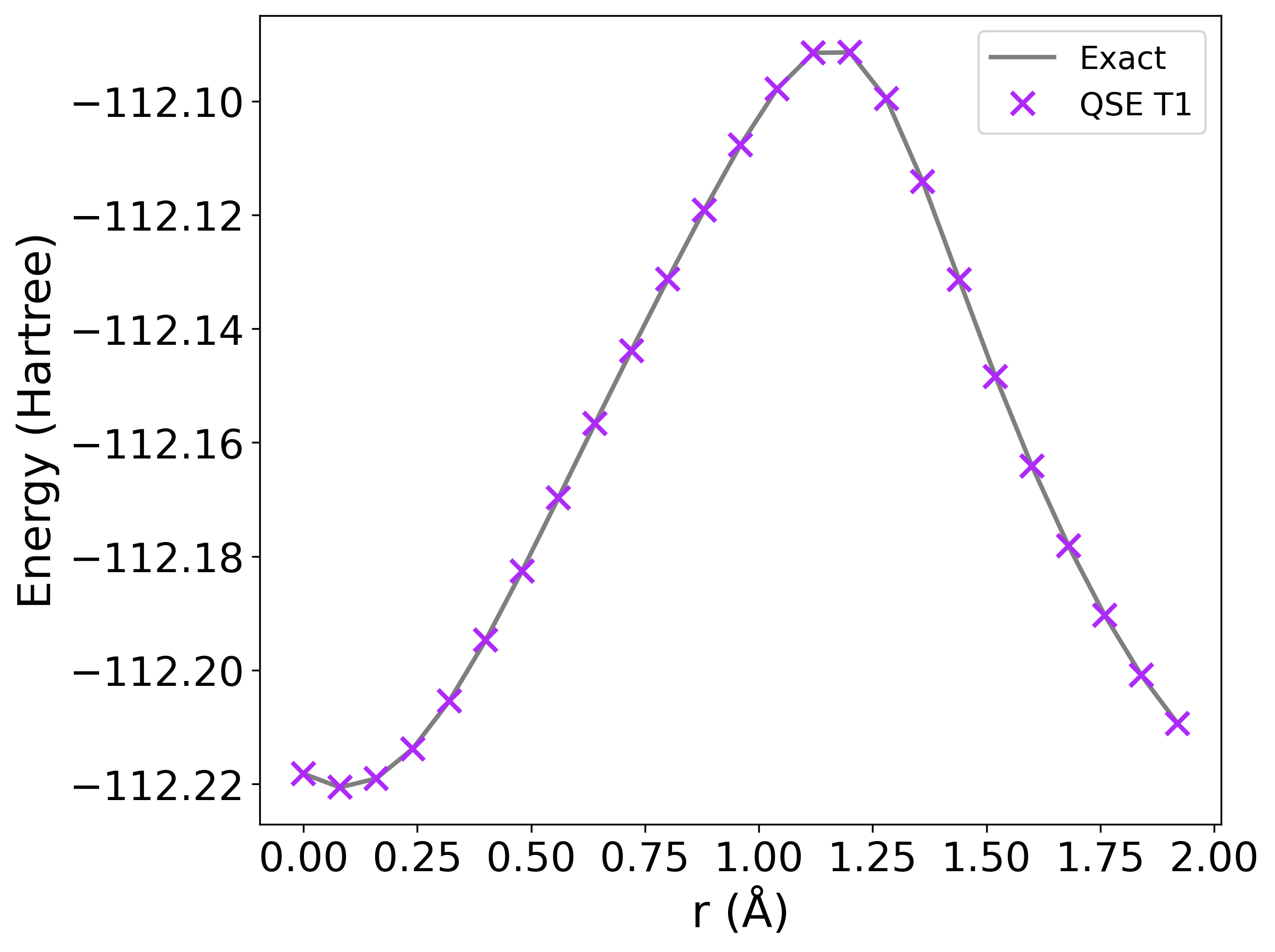}
\caption{}
  \end{subfigure}

\caption{Triplet PES results by QSE (a)CAS(3,2) T$_1$ and T$_2$ of H$_3^+$ (b)CAS(3,2) T$_1$ of Ch$_2$O.}
  \label{triplet}
\end{figure}

\begin{table}[h]
\centering
\caption{Comparison of QSE T1 and T2 $\Delta$E Errors (Hartree) for H$_3^+$ and CH$_2$O.}
\begin{tabular}{lcccc}
\toprule
$\Delta$E (Hartree) & T1 Average & T2 Average & T1 Middle & T2 Middle \\
\midrule
\multicolumn{5}{c}{H$_3^+$} \\
RMSE & \num{1.14e-03} & \num{1.83e-04} & \num{0.0} & \num{0.0} \\
Max Error & \num{5.65e-03} & \num{9.99e-04} & \num{0.0} & \num{0.0} \\
MAE & \num{4.19e-04} & \num{3.75e-05} & \num{0.0} & \num{0.0} \\
\midrule
\multicolumn{5}{c}{CH$_2$O} \\
RMSE & \num{6.80e-08} & - & \num{0.0} & - \\
Max Error & \num{3.40e-07} & - & \num{0.0} & - \\
MAE & \num{1.36e-08} & - & \num{0.0} & - \\
\bottomrule
\end{tabular}
\label{triplettab}
\end{table}
To further explore the utility of the subspace quantum\added{-computing electronic structure} solver, we evaluated its performance in computing triplet state energies in Figure~\ref{triplet} and Table \ref{triplettab}. For CH$_2$O, the CAS(3,2) active space consisted of the three Hartree–Fock canonical orbitals straddling the Fermi level. Here, the QSE consistently yields three degenerate or near-degenerate energy values for triplet states. We explored two strategies for processing these values: averaging the three energies or selecting the median value, with the latter often proving more accurate. These findings demonstrate QSE's potential for investigating more sophisticated excited-state dynamics in future studies.

\clearpage
\subsection{Computational Resource Analysis}
\label{resource_analysis}
\begin{figure}[]
  \centering

  \begin{subfigure}[b]{0.80\textwidth}
    \centering
    \includegraphics[width=\textwidth]{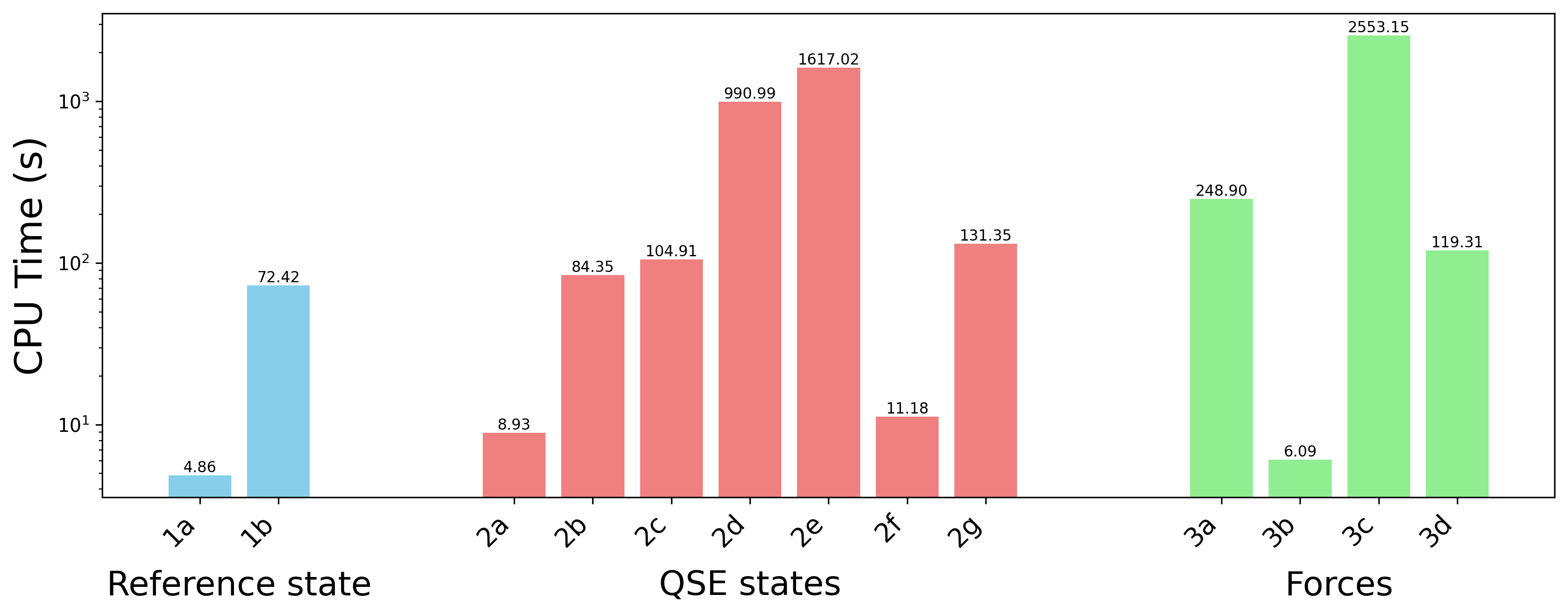}
    \label{efficiency}
\caption{}
  \end{subfigure}
  \hfill
  \begin{subfigure}[b]{0.19\textwidth}
    \includegraphics[width=\textwidth]{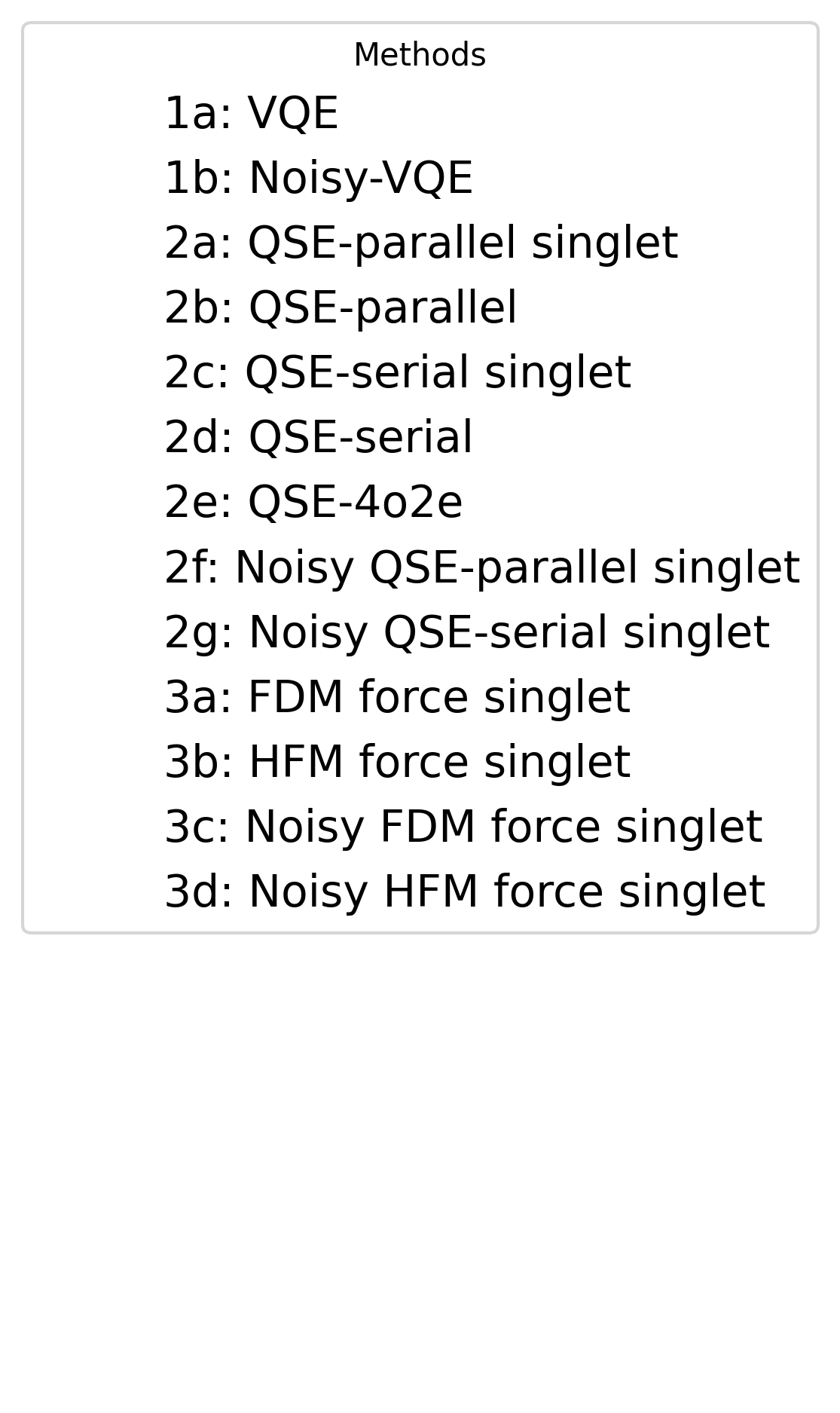}
    \caption{}
  \end{subfigure}

  \vspace{0.5cm}
  \begin{subfigure}[b]{0.98\textwidth}
    \centering
    \includegraphics[width=\textwidth]{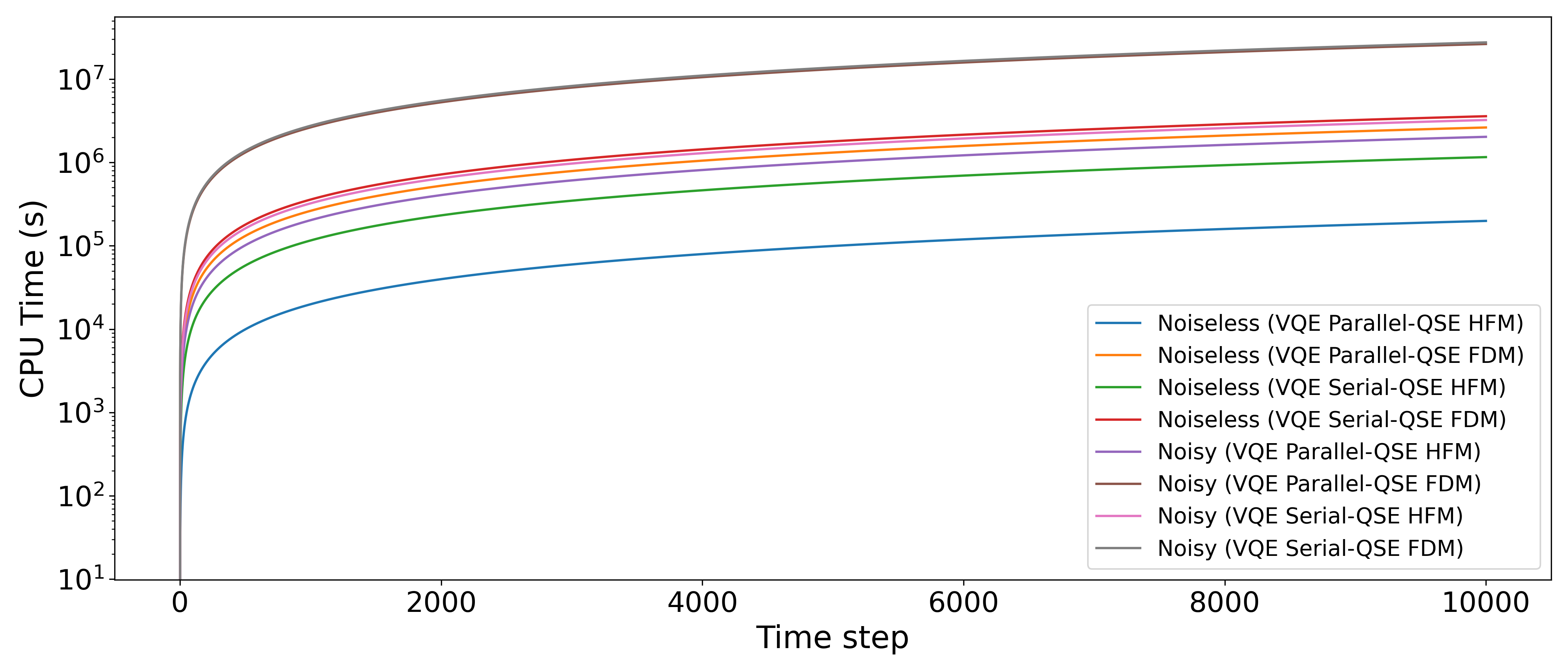}
\caption{}

  \end{subfigure}

  \caption{CPU time comparison and estimation of the simulation. Except for 2e cases in plot (a), all the QSE implementations are tested on CAS(3,2). (a) CPU time of different QSE implementations on calculating energies and forces. (b) Legend for plot (a). (c)CPU time estimation for classical-computing simulation of different quanutm-computing LZSH parallel strategy.}
  \label{fig:cputimee} 

\end{figure}
To quantify the impact of our parallelization strategies on computational efficiency, we present benchmark results from \replaced{executing QSE algorithm via quantum algorithm simulator on classical computer}{classical emulations of the QSE algorithm}, focusing on wall-clock times for key components of the excited-state dynamics simulation. Figure~\ref{fig:cputimee} (a) illustrates the CPU times required for ground-state energy evaluations using VQE with UCCSD (1a: 4.86 s) and its noisy variant (1b: 72.42 s), excited-state subspace diagonalizations via various QSE implementations (2a-2g), and force computations employing FDM or HFM approaches (3a-3d). Notably, the parallelized QSE implementations demonstrate substantial speedups over their serial counterparts. For the spin-adapted singlet variant, the parallel QSE (2a: 8.93 s) achieves approximately a 12-fold reduction compared to the serial QSE (2c: 104.91 s), attributable to the concurrent evaluation of expectation values for the independent Pauli strings in the $H$ and $S$ matrices (Eqs.~\eqref{eq:qse_matrices}). Similarly, for the triplet QSE, parallelization (2b: 84.35 s) yields a comparable ~12× speedup relative to the serial case (2d: 990.99 s). The CAS(4,2) (4o2e) QSE-triplet (2e: 1617.02 s), which incorporates excitatoin operators, incurs additional overhead due to the increased subspace dimension, highlighting the trade-off between accuracy and efficiency in larger subspace. Under noisy conditions, emulating realistic quantum hardware errors, the parallel singlet QSE (2f: 131.35 s) remains efficient. For QSE*, since we have expanded more operators and conducted parallel estimation on these operators, the acceleration effect will be more significant if there is sufficient hardware.

At the force computation level, the HFM approach (3b: 6.09 s) outperforms FDM (3a: 248.90 s) by ~41× in the noiseless regime, reflecting the lower measurement demands of direct derivative evaluations compared to FDM. Noise channel simulation amplifies this disparity, with noisy FDM (3c: 2553.15 s) being ~21× slower than noisy HFT (3d: 119.31 s).

Extending to full trajectory propagation, Figure~\ref{fig:cputimee} (c) depicts the cumulative computation time as a function of time steps for a representative non-adiabatic dynamics simulation under LZSH, comparing noiseless and noisy scenarios with parallel versus serial QSE integrated into VQE for ground-state preparation, alongside HFT or FDM for forces. The parallel QSE configurations consistently exhibit shallower slopes, indicating reduced per-step overhead. For instance, the noiseless VQE with parallel-QSE and HFT (blue line) accumulates ~$10^5$ s by 10,000 steps, whereas the serial-QSE equivalent (green line) approaches ~$10^6$ s, a ~10× difference arising from the distributed measurement strategy. This gap widens in noisy emulations, where parallel-QSE with FDM (purple line) remains below $10^7$ s, while serial-QSE with FDM (magenta line) exceeds it, emphasizing the robustness of parallelization to noise channal computation. The FDM variants generally worth than HFT counterparts within each category, aligning with the single-point benchmarks, though the asymptotic scaling remains dominated by the QSE matrix construction.

Beyond the per-trajectory level, our trajectory-level parallelization exploits the parallel nature of the ensemble, distributing independent Wigner-sampled trajectories across computational nodes. While not explicitly benchmarked here due to hardware constraints, scaling analyses suggest near-linear speedups with the number of processors, limited only by load balancing in asynchronous hop events. Collectively, these optimizations reduce overall simulation wall times by 1-2 orders of magnitude for typical photochemistry applications with 10-100 trajectories and active spaces of 4-8 orbitals, paving the way for efficient quantum-classical hybrid simulations.

\added{While our benchmarks focus on validating the efficiency gains from parallelization (via classical-computing emulations of quantum algorithms), it is important to contextualize these results against traditional classical computational chemistry software. At present, simulations of quantum algorithms on classical computer are substantially slower than well-optimized classical programs (e.g., those implemented in C++/Fortran with extensive algorithmic refinements), stemming from the emulation's need to mimic quantum operations through matrix-vector computations (or matrix-matrix computation for noisy quantum circuit) with limited optimizations to preserve universality. Moreover, current quantum hardware lacks the necessary fidelity for high-precision electronic structure calculations, precluding direct comparisons of real quantum runtimes.}

\added{In terms of user experience, classical softwares (e.g. Pyscf\cite{pyscf}, Molpro\cite{molpro}, Psi4\cite{psi4}, etc.) offer seamless access to a broad array of properties beyond energies, such as spin expectation values (\(S^2\)) and configuration interaction coefficients facilitated by storing the full wavefunction in classical memory for efficient post-processing. Quantum approaches, by contrast, require explicit measurements for each desired observable---e.g., additional shots for \(S^2\) via spin operators---and accessing configuration interaction coefficients demands quantum state tomography to retrieve all \(2^n\) amplitudes ($n$ be the number of spin orbitals), which is exponentially challenging and resource-intensive. Thus, while quantum methods hold promise for scaling to larger systems, classical quantum chemistry software currently provide a more convenient and comprehensive workflow.}\comment{R3 Q19}

\clearpage
\section{Conclusion}
In this work, we have developed a efficient quantum computational framework for NAMD, in which, parallelization is supported both for high-precision PES calculation and quantum-algorithm adapted LZSH trajectory simulation. Our approach integrates the CAS framework with VQE and its subspace variant SSVQE, to adaptably prepare reference states. Additionally, we incorporate QSE and its extended variant QSE* for accurate excited-state calculations. Beyond energy spectrum computation, our method enables the calculation of nuclear forces by interfacing quantum\added{-computing PES} solvers with the HFM and FDM. Another advancement is the seamless integration of quantum algorithms with the LZSH framework, augmented by curvature-induced hopping corrections to mitigate PES fluctuations at dissociation limits.

Numerical benchmarks on H$3^+$, C$_2$H$_4$, and CH$_2$O demonstrate sub-microhartree accuracy on PESs by our hybrid subspace quantum\added{-computing electronic structure} solvers. By validating problem-tailored QSE operator extension, we enhance the adaptability of our approach across those diverse chemical systems. The incorporation of quantum\added{-computing electronic structure} solvers and curvature-driven hopping correction in LZSH significantly improves the robustness of NAMD simulations while preserve efficiency, as evidenced by comparisons with \deleted{classical} exact \added{reference} results. Furthermore, computational resource analysis show that our two-level parallelization framework delivers substantial computational speedups, fully transferable to \added{real} quantum \replaced{computer}{hardware}, without compromising precision.

This work advances quantum computational NAMD by addressing critical bottlenecks in efficiency and robustness at both trajectory level and electronic structure levels, while enhancing the adaptability and precision of PES calculation. These advancements pave the way for exploring non-adiabatic effects in polyatomic molecules beyond classical computational limits while facilitating the practical utility of quantum computing. However, challenges persist, including the systematic handling of $S$ matrix illness in QSE, ansatz expressivity in VQE, and the integration with advanced infrastructures to handle quantum noise. Future efforts will focus on integrating orbital optimization techniques, such as complete active space self consistent field method (CASSCF), developing systematic QSE extension strategies, and exploring the embedding of quantum computing on other NAMD frameworks.

\section{Data availability}
The data supporting this article have been included as part of the Supplementary Information of this article. Data and scripts for this paper, including the PES, forces and NAMD initial conditions, are available at Zenodo at: \url{https://doi.org/10.5281/zenodo.17865112}. The code utilized in this study is available at in the project Gitee repository: \url{https://gitee.com/mindspore/mindquantum/new/research/paper_with_code/Efficient_Quantum_Simulation_of_Non-Adiabatic_Molecular_Dynamics_with_Precise_Electronic_Structure}.

\begin{acknowledgement}
This work is supported by the NSAF (Grant No. U2330201), National Natural Science Foundation of China (Grant No. 22303005), Innovation Program for Quantum Science and Technology (Grant No. 2023ZD0300200), the CPS-Yangtze Delta Region Industrial Innovation Center of Quantum and Information Technology-MindSpore Quantum Open Fund, and the High-performance Computing Platform of Peking University.

\end{acknowledgement}

\begin{suppinfo}
The data supporting this article available. See DOI: \url{https://doi.org/10.5281/zenodo.17865112}.
\end{suppinfo}

\bibliography{achemso-demo_1}

\end{document}